# Whispering gallery microsensors: a review

Xuefeng Jiang, Abraham J. Qavi, Steven H. Huang, and Lan Yang*

*Department of Electrical and System Engineering, Washington University in St. Louis, St. Louis, Missouri 63130, USA*

**Corresponding author: yang@seas.wustl.edu*

## Abstract:

*Optical whispering gallery mode (WGM) microresonators, confining resonant photons in a microscale resonator for long periods of time, could strongly enhance light-matter interaction, making it an ideal platform for all kinds of sensors. In this paper, an overview of optical sensors based on WGM microresonators is comprehensively summarized. First, the fundamental sensing mechanisms as well as several recently developed enhanced sensing techniques are introduced. Then, different types of WGM structures for sensing as well as microfluidics techniques are summarized. Furthermore, several important sensing parameters are discussed. Most importantly, a variety of WGM sensing applications are reviewed, including both traditional matter sensing and field sensing. Last, we give a brief summary and perspective of the WGM sensors and their role in future applications.*

## I. Introduction

Optical sensors have become one of the elementary building blocks of modern society, in which we rely on them to facilitate our everyday lives by monitoring the intensity, phase, polarization, frequency or speed of the light. A prominent example is fiber optic sensor [1]–[3], which have been well developed to measure strain, temperature, pressure, voltage or other physical quantities in an optical fiber as well as fiber-based devices, and particularly to provide distributed sensing capabilities over a very long distance. For single-pass fiber optic sensors, to improve the sensitivity as well as the detection limit, one typically needs to increase the physical length of the fiber to enhance the interaction between the light and target matter/field. On the other hand, resonant micro/nano structures provide an excellent class of optical sensors. These structures allow for high quality factors ($Q$) of the resonant system while keeping the device compact, enabling resonant optical fields to travel inside the resonator for multiple times, and thus dramatically improving the sensitivity of the devices. While there are many types of microresonator structures, whispering gallery mode (WGM) microresonators possess the highest quality factor ($Q$). This property has catapulted WGM microresonators to the forefront of photonic sensors, and is demonstrated by the rich scientific community surrounding these devices over the past two decades. [4]–[18]

WGM based sensors were first developed by Vollmer and Arnold in 2002 for the detection of proteins in aqueous environments by monitoring the resonant frequency shift in a WGM microsphere [19]. This research field of WGM biosensors has seen enormous growth since then, and has been summarized by several review articles [4]–[18]. In addition to biomoecules, WGM sensors have been employed for chemical sensing, temperature sensing, gas sensing, electric/magnetic field sensing, and pressure/force sensing. Furthermore, several novel sensing mechanisms, including exceptional point enhancement, cavity ring-up, backscattering reflection, and optomechanics have been developed in the past few years, and have yet to be well reviewed.

In this paper, we review the mechanisms, methods, structures, technique, parameters, and applications of WGM sensors. Sections II and III briefly discuss three fundamental sensing mechanisms of WGM microresonators, as well as several recently developed enhanced sensing techniques. In Section IV, different kinds of WGM structures as the platforms for sensing as well as microfluidics techniques are presented. Section V summarizes some important sensing parameters. Most importantly, a variety of WGM sensing applications are presented in Section VI, including not only traditional matter sensing, such as particle, gas and bio/chemical sensing, but also field sensing, such as temperature, electric/magnetic field, and pressure/force sensing. Finally, in Section VII, we briefly summarize the field as well as future directions of WGM sensors.

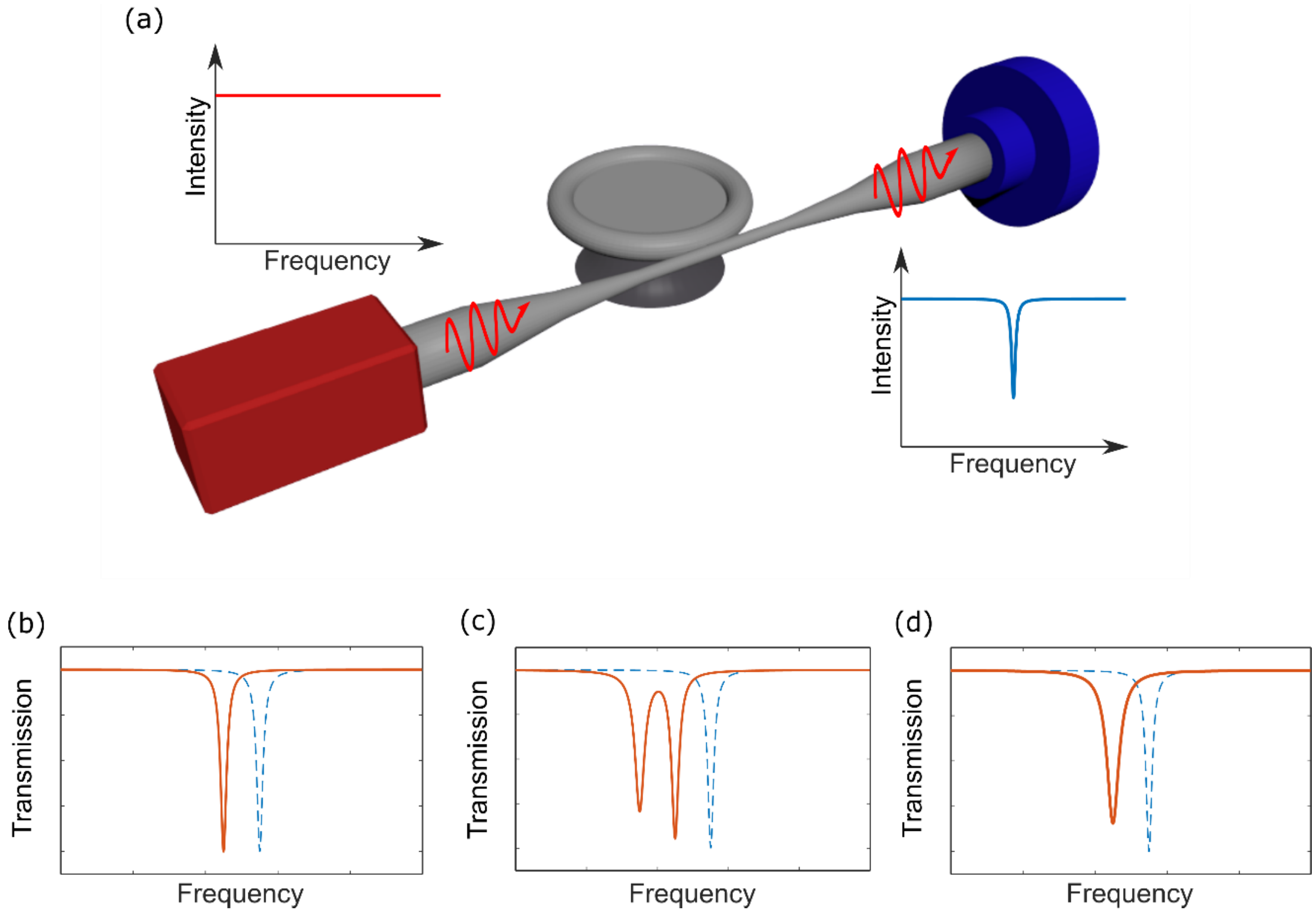

Fig. 1. Concept (a) and basic sensing mechanisms (b-d) of the WGM sensor, including mode shift (b), mode splitting (c) and mode broadening (d).

## II. Fundamental sensing mechanisms

### A. Mode shift

In mode shift sensing, changes in the resonant wavelength of WGMs are used to measure the signal of interest [9]. These changes are most often measured directly through the resonator's transmission, reflection, or emission spectra, but can also be measured indirectly by locking the probe wavelength and observing changes in the transmission instead. Mode shift is the most commonly used sensing modality for WGM resonators, due to its broad applicability to a range of analytes. Mode shift sensing can also be used to measure the adsorption of analytes such as single nanoparticles [9] and single molecules, [20] and thin layers of material. [19] It can also be used to measure bulk refractive index changes surrounding the microresonator, such as that caused by varying the amount of ethanol in water. [21] Mode shift sensing is also used to detect changes in physical parameters surrounding the WGM device [8], such as heat, pressure, and magnetic fields. Although this makes the mode shift technique incredibly versatile, it is also problematic as changes in ambient conditions (such as temperature, pressure, and humidity) tend to introduce unwanted drift in the mode shift signal.

The mode shift observed in WGM resonators due to the adsorption of analyte can be understood by the reactive sensing principle. Intuitively, when a particle with refractive index greater than the medium around the resonator is adsorbed on the resonator, it pulls a part of the optical field outward, increasing the optical path length and leading to a red-shift in the resonance mode. More precisely, by first order perturbation theory, the amount of frequency shift, $\delta\omega_0$ induced by a single nanoparticle/molecule is expressed by: [9], [22]

$$\frac{\delta\omega_0}{\omega_0} = \frac{-\alpha_{ex}|\boldsymbol{E}_0(\boldsymbol{r_i})|^2}{2\int \epsilon_r(\boldsymbol{r})|\boldsymbol{E_0}(\boldsymbol{r})|^2 dV}$$

where $\alpha_{ex}$ is the excess polarizability of the particle, $\boldsymbol{r_i}$ is the position of the particle, and $\epsilon_r$ is the permittivity of the medium. Note the mode shift's dependence on the position of the particle; this makes it challenging to correlate the amount of shift to particle size for a single particle measurement. Given this limitation, mode shift sensing requires finding the maximum shift from a large number of single particle events, to be able to develop this correlation. [9]

### B. Mode splitting

Mode splitting relies on the measurement of scatterer-induced coupling between the clockwise (CW) and counterclockwise (CCW) propagating WGMs. This mechanism is used almost exclusively for sensing the adsorption of nanoparticles on resonators and it works the best for sensing nanoparticles on the order of 10-100 nm. [23]–[35] To explain this phenomenon, first note that in a WGM resonator without scatterer, resonant modes exist in pairs of degenerate modes: the CW propagating modes and CCW propagating modes. When there is a scatterer on the resonator, part of the light scattered from the CW mode is scattered into the CCW mode, and vice versa, inducing coupling between the two modes. [36], [37] Due to this coupling, the degeneracy between the two modes are lifted. By measuring this resulting resonant frequency splitting, the presence of the scatterer can be detected.

One of the advantages of the mode splitting mechanism is its ability to perform self-referenced detection, as the two split modes serve as a reference to each other. As a result, mode splitting is not influenced by thermal or pressure fluctuations that are present in mode shift. Another advantage is that by observing the change in the transmission spectrum with mode splitting, it is possible to determine the size of particles adsorbed to the resonator surface in real-time. [23], [38]

C. Mode Broadening

Another sensing mechanism is resonant linewidth broadening [39], which combines the advantages of both mode shift and mode splitting. Mode linewidth broadening is used to measure either scattering loss [39], [40] or absorption loss caused by the adsorption of analytes such as single nanoparticles [41] and single biomolecules, and bulk refractive index change in the surrounding [42], [43]. Similar with mode splitting, it is also a self-referenced detection mechanism, isolating the fluctuation of environment temperature or system instability. Furthermore, it removes the requirement of narrow linewidth (ultrahigh *Q*) needed to resolve the doublet in the splitting spectrum. Experimentally, Shao *et al.* reported the detection and counting of single 70 nm-radius polystyrene (PS) nanoparticles and lentiviruses by monitoring the linewidth broadening in a free-space coupled deformed microresonator, both of which were verified via SEM imaging. [39] The theoretical detection limit for single PS nanoparticle using the mode broadening measurement is as small as 10 nm in radius. On the other hand, Armani and Vahala demonstrated the detection of the heavy water with a concentrations of 0.0001% $D_2O$ in $H_2O$ by monitoring the absorption induced linewidth narrowing (*i.e., Q* factor increasing) [42]. The mode broadening mechanism also holds great potential for gas sensing by monitoring the gas absorption in the future.

## III. Enhancement methods and new techniques

A. Plasmonic Enhancement

When light is incident on gold or silver nanoparticles, the free electrons in the metallic nanoparticles oscillate with the incident optical field, forming localized surface plasmon (LSP) resonance. LSP resonance creates regions around the nanoparticles with intense electromagnetic field, often called plasmonic hotspots. [20], [44]–[51] When a molecule of interest enters one of these plasmonic hotspots it interacts strongly with the plasmonic resonance, rendering the plasmonic resonance a highly sensitive technique for detecting surface adsorbed layers. LSP resonance can be used in combination with WGM resonance to achieve single molecules detection. An excellent review has recently appeared on hybrid plasmonic-photonic WGM resonators, [52] and the readers are referred to this review for more details.

Early works on hybrid plasmonic WGM resonator for sensing demonstrated the detection of single PS particles, single virus particles, and proteins in solution with enhanced sensitivity. [46], [53]–[55] The first single-protein detection using WGM resonator was demonstrated by Dantham *et al.* using gold nanoshells. [20] The authors noted that the plasmonic enhancement from a smooth nanoshell alone was not able to explain the observed enhancement in sensitivity, and rather the large enhancement in sensitivity was attributed to bumps on the nanoshell that enhanced the local field strength even further. Later, Baaske *et al.* demonstrated the measurement of single nucleic acids and even single atomic ions using a hybrid gold nanorod-WGM resonator, marking the highest sensitivity achieved by WGM resonators so far. [47], [48] In these works, the detection of single molecules appeared as either spikes or step changes, corresponding to transient or semi-permanent binding events. The statistics of this type of binding events were shown to correspond to the binding affinities of molecules, [56] and such measurement can provide a powerful tool to study molecular interactions on surfaces.

In a slightly different use of plasmonic particles in combination with WGM resonators, Heylman *et al.* demonstrated the absorption spectroscopy and photothermal mapping of gold nanorods on microtoroid resonators. [57] In this work, the authors also noted evidence of WGM-plasmon interactions in the form of Fano resonances, which may result in further enhancement in sensitivity for molecular binding events in the future. [58]–[60]

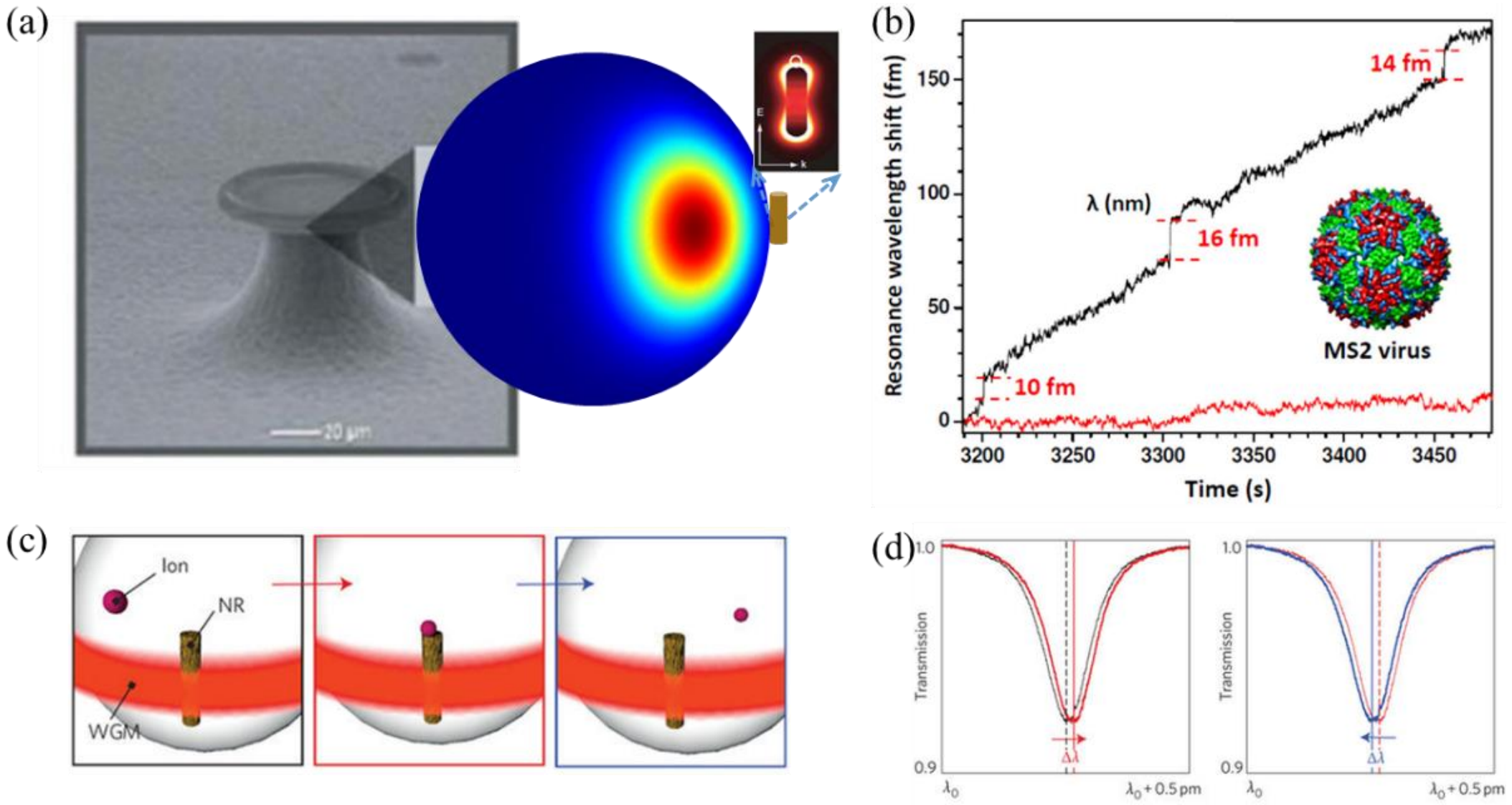

Fig. 2. Plasmonic enhanced WGM sensing. (a) An SEM image of a microtoroid coupled with an equatorially bound Au nanorod (NR). Insets: Electric field intensity distributions of a nanorod and WGM. (b) Resonance wavelength shift when individual $MS_2$ viruses bind to the microsphere with a gold nanoshell attaching on its equator. The control experiment, *i.e.*, without the nanoshell or the MS2 virus, is shown by the red curve. (c), (d) Plasmonic enhanced transient sensing. Transient interactions of single ions with the NRs (c), excited at their plasmon resonance, are detected as a red shift of the WGM resonance (d, left) when an ion enters a sensing site on a NR's surface, and a subsequent blue shift (d, right) when the ion leaves the sensing site.

## B. Self-Heterodyned Microlasing

The detection limit of a WGM resonator, when using the mode splitting mechanism, depends on the linewidth of the WGM as the linewidth is related to the resolvability of the split modes. This limit of detection can be enhanced enormously by observing the laser spectrum in an active cavity rather than the transmission spectrum of a passive cavity, as the laser linewidth is much narrower than the corresponding passive cavity linewidth [61]–[73]. Experimentally, the gain in the silica microcavity can be provided by either rare earth ion doping [24], [74], [75], such as erbium, yttrium, thulium, or via stimulated Raman scattering operating at any wavelength. [28], [29], [76]–[79] The direct measurement of mode splitting spectrum through the lasing spectra is typically not possible since the laser splitting is smaller than the resolution of optical spectrometers. Instead, the mode splitting is measured through observing the beating in the laser output. [24], [80] As shown in Fig. 3, when a nanoparticle/molecule moves into the mode volume of the WGM, the laser spectrum splits. The split laser modes lead to a beat note with a frequency that is equal to the difference in frequency between the two laser modes. Furthermore, the lasing spectrum and the frequency of the beat note change again when a second nanoparticle binds. Accordingly, individual nanoparticles can be continuously detected in real time by monitoring the beat note signal.

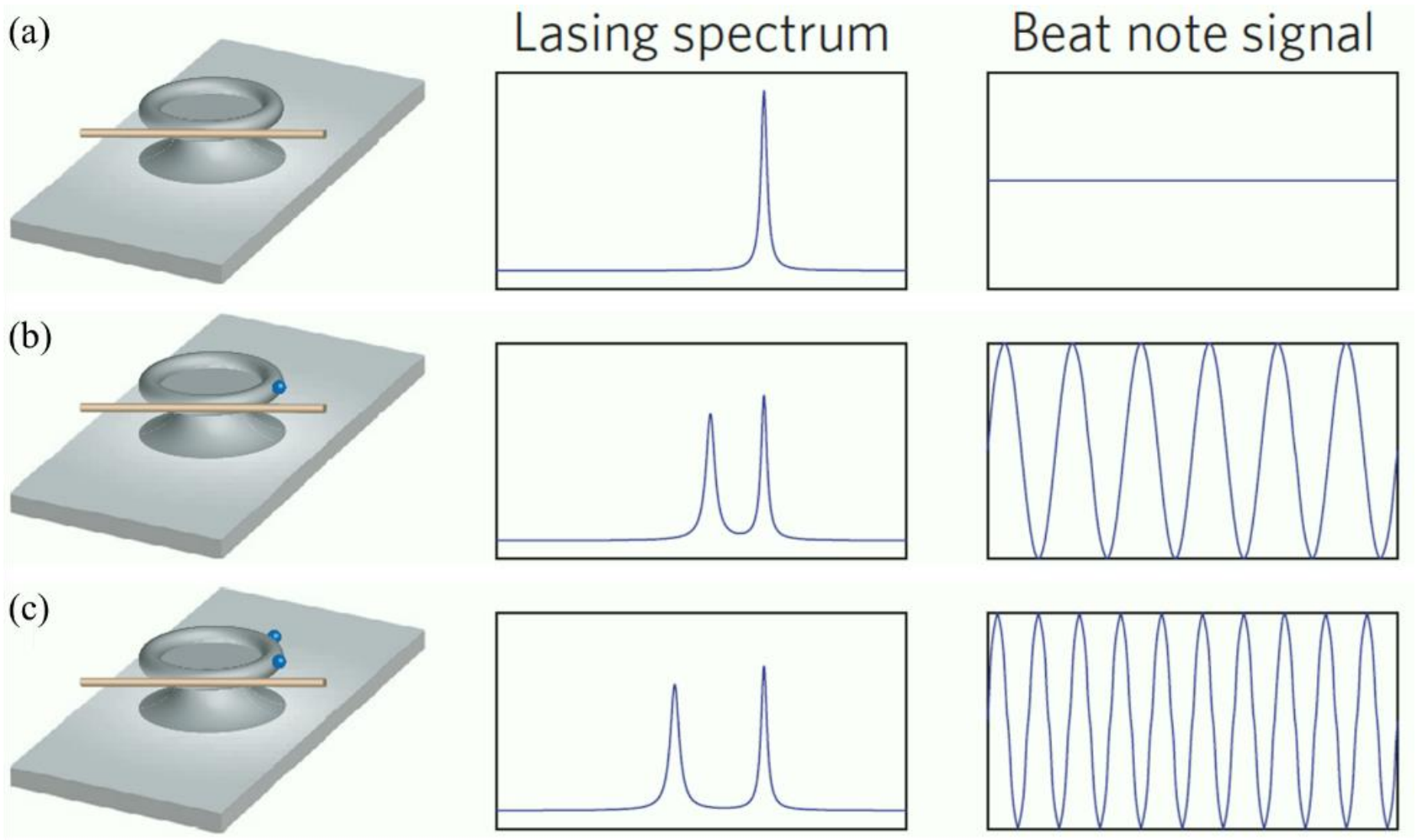

Fig. 3. Self-heterodyned detection of single nano-objects using microlasing. (a) Before nanoparticles arrive there is a single laser mode and the laser intensity is constant. (b) The lasing mode splits into two when the first nanoparticle binds, leading to a beat note with a frequency that is equal to the difference in frequency between the two modes. (c) The lasing spectrum and the frequency of the beat note change again when a second nanoparticle binds.

## C. Exceptional Point

Optical cavities operating at non-Hermitian spectral degeneracies known as exceptional points (EPs) have been shown to demonstrate non-trivial physical phenomena, such as non-conventional dependence of laser power on pump and loss. [81], [82] It has been recently demonstrated that at such exceptional points, the mode-splitting induced by a scatterer can be enhanced, as shown in Fig. 4. [83], [84] The enhancement comes from the square-root dependence of mode-splitting on particle induced perturbation near second-order EP, as opposed to the linear dependence in conventional cavity sensors. This enhancement in sensitivity is greater for smaller perturbations and a factor of 2 enhancement in sensitivity has been experimentally demonstrated, by tuning WGM to EP with two nanotips. [83] Single 200-nm PS nanoparticles have also been detected with

EP enhancement.

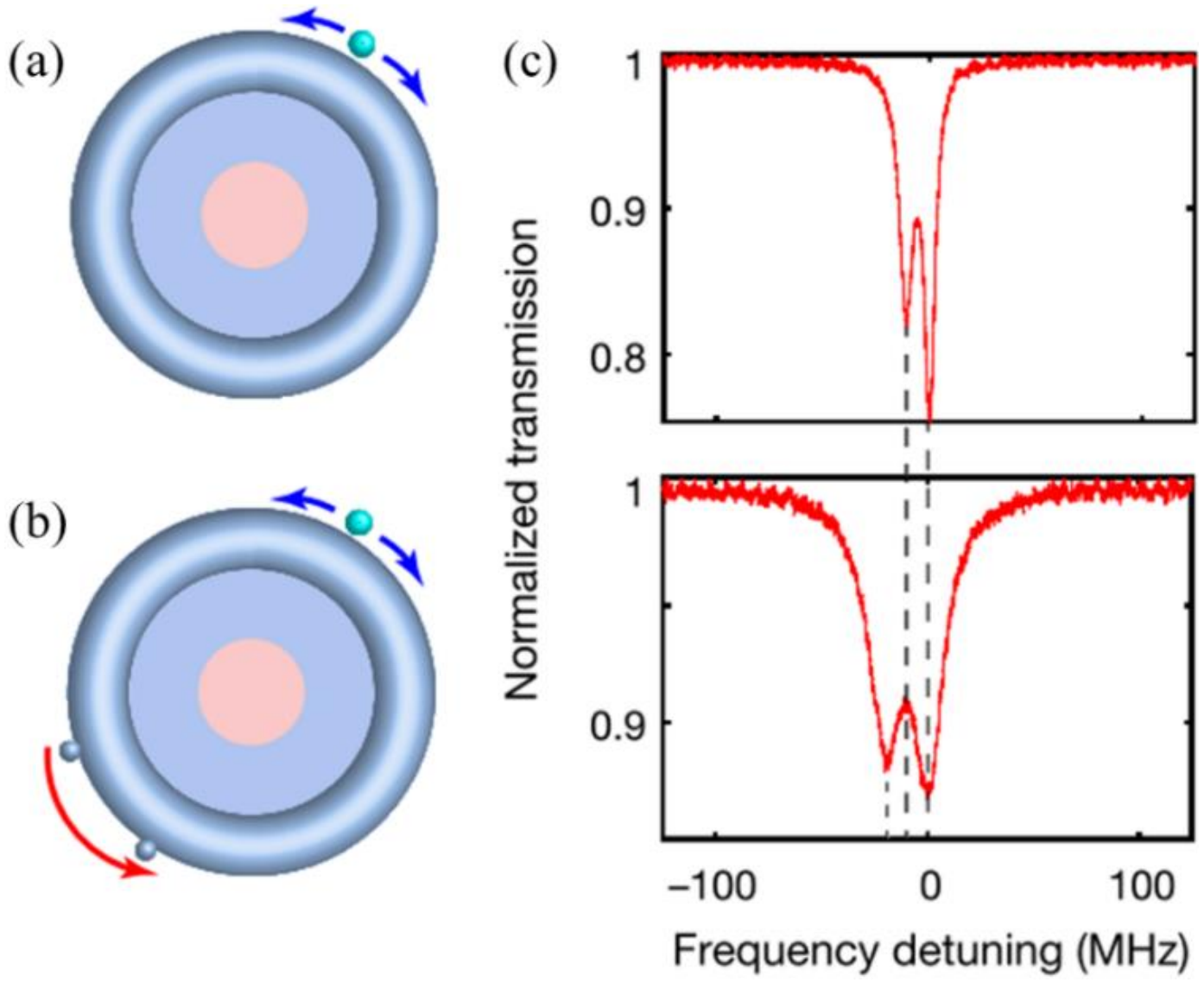

Fig. 4. Mode splitting sensitivity enhancement at exceptional points. Diagrams of (a) traditional and (b) EP enhanced WGM sensors and (c) corresponding transmissions. The blue arrows illustrate the symmetric backscattering of the target particle and the red arrow designates the fully asymmetric backscattering related to the exceptional point. [84]

## D. Mode Locking

Traditional sensing mechanisms, such as mode shift/splitting/broadening, require frequency scanning of the probe laser to obtain the mode spectrum of the microcavity. The acquisition time is usually on the order of ten milliseconds, limited by the frequency modulation bandwidth of the laser (typically <2 kHz). For mode shift sensing, an alternate method to improve time resolution is the mode locking technique [85], [86]. Specifically, the probe laser frequency is locked to the resonant frequency of a WGM by the Pound–Drever–Hall (PDH) technique, as shown in Fig. 5. A small high-frequency dither is utilized to modulate the probe laser frequency, which is multiplied with the transmitted spectrum. After time averaging, an error signal is generated, and its amplitude is proportional to the frequency difference of the current probe laser and resonant mode. The resonant mode shift signal induced by either single nanoparticle/molecule binding or environment field/refractive index changing could be extracted from the feedback error signal. The time resolution of the mode locking system is as small as 1.2 ms, and single 39 nm ×10 nm gold nanorods have been detected using this technique [85]. Furthermore, Su *et al.* demonstrated that the mode locking technique possessed capacity of single nanoparticles/biomolecules detection with radii ranging from 2 to 100 nm in an aqueous environment by optimizing the detection system and post-processing filtering routines (Fig. 5(c)). [86]

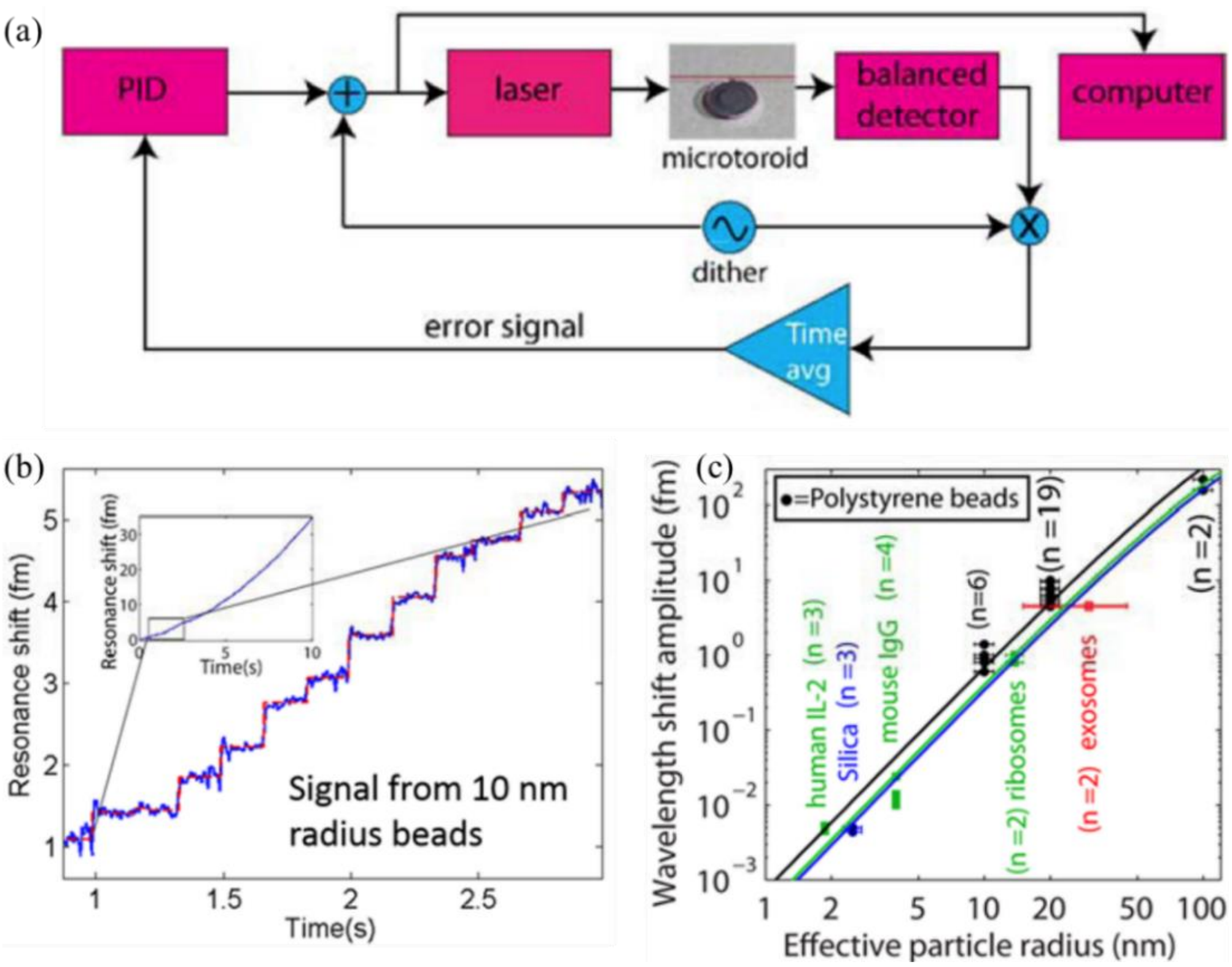

Fig. 5. (a) Diagram of a mode locking sensing system. (b) The mode shift signal over time as 10-nm-radius polystyrene (PS) nanoparticles bind to the surface of the microtoroid. Inset: Sensing response over the full recording range of 10 seconds. (c) Summary of single particle/molecule detection data based on mode locking technique with radii from 2 to100 nm. [86]

## E. Ring-Up Spectroscopy

Although sub-millisecond time resolution can be achieved with the mode locking technique, it can only be used to measure the resonant frequency shift of a WGM. Cavity ring-up spectroscopy, on the other hand, provides a solution for measuring the mode shift, splitting, and broadening signals synchronously [87]–[89]. This technique offers a time resolution as short as 16 ns per frame [87]. Specifically, a blue-detuned probe laser pulses is coupled with the modes, resulting in a build-up transient field in the cavity, which interferes with the transmitted field to create a ring-up signal, as shown in the center inset of Fig. 6. In the ring-up signal, the detuning, $\delta$, from the probe is exhibited by the fast oscillations; the resonance width, $2\kappa$, is exhibited by the exponential decay envelope; and the slow beat note indicates the splitting of the resonance, $2g$. A time resolution of 16 ns can be obtained by measuring the optomechanical oscillations. Cavity ring-up spectroscopy provides an opportunity for optical WGM sensing with nanosecond time resolution and holds great potential to detect single nanoparticle/molecule movement by deriving the signals of the mode shift, splitting, and broadening from the ring-up signal.

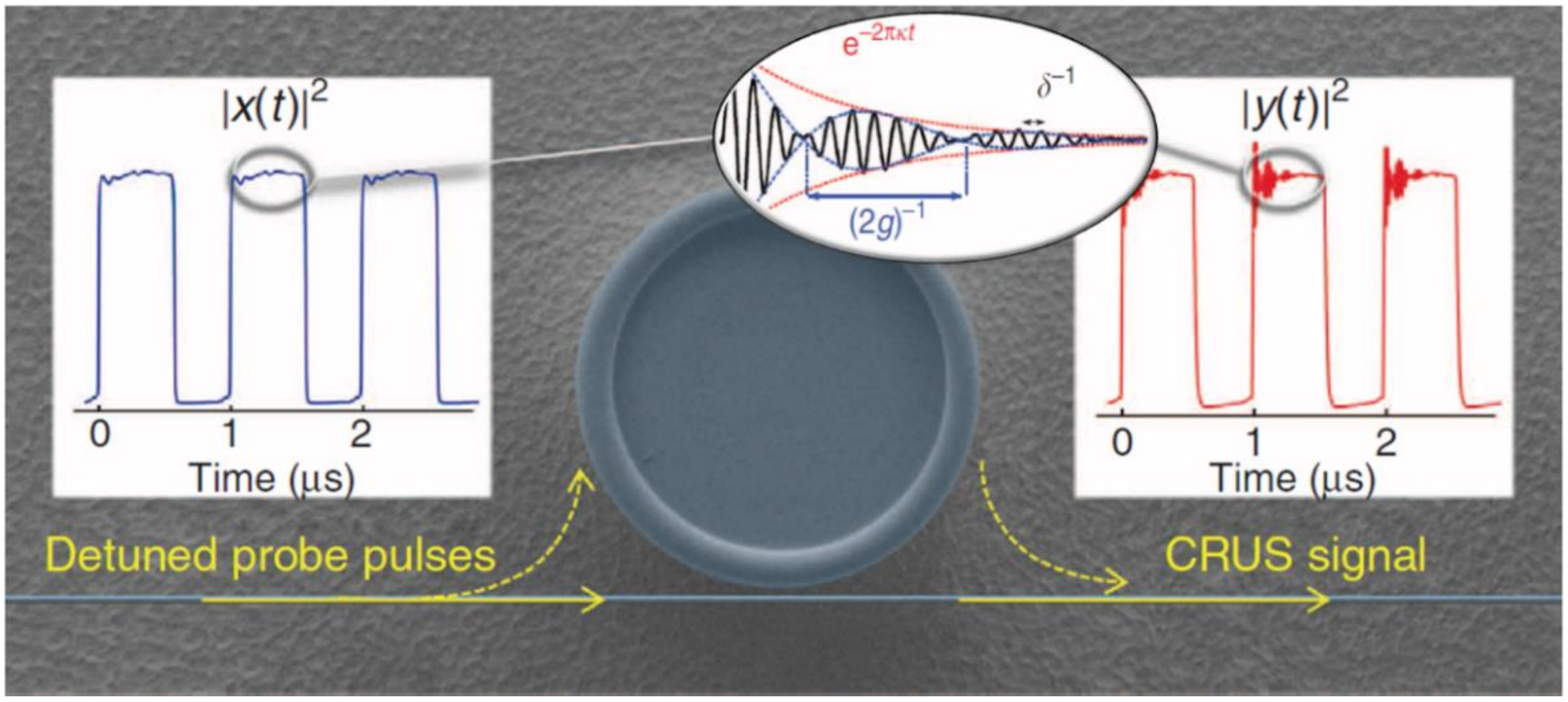

Fig. 6. SEM image of a tapered 1fibre-coupled microtoroid. Sharply rising detuned probe pulses (left inset) lead to the build-up of a cavity field. As this weak field leaks back into the fiber, it interferes with the probe, resulting in a beating signal at the output (right inset). The resonance width, $2\kappa$, the detuning, $\delta$, and the splitting of the resonance, $2g$, can be derived from the ring-up signal, showing in the center inset. [87]

## F. Backscattering Reflection

The above sensing mechanisms/techniques are all based on the optical transmission spectrum. Another sensing method is to observe the reflection spectrum of the back-scattered field induced by the target nanoparticles/molecules, which has been achieved in WGM microresonators. [90], [91] Experimentally, single 50 nm PS particles and 20 nm sodium chloride (NaCl) particles were detected by monitoring the reflection intensity in a microcavity-taper system (Fig. 7). [91] One of the advantages of the reflection sensing is the lower noise level comparing with the transmission signal, which will result in an improved detection limit. The experimental noise of the reflection measurement has been discussed by Knittel *et al.* [90], and is shown in the inset of Fig. 7(b). Note that the noise level of the back-scattering spectrum is lower than that of PDH spectrum over the full frequency window. The back-scattering noise level is dominated by laser intensity induced $1/f^2$ noise, which can be greatly suppressed by either employing a noise eater in the system, or measuring noise simultaneously by the back-scattering system and then subtracting it in the post-processing.

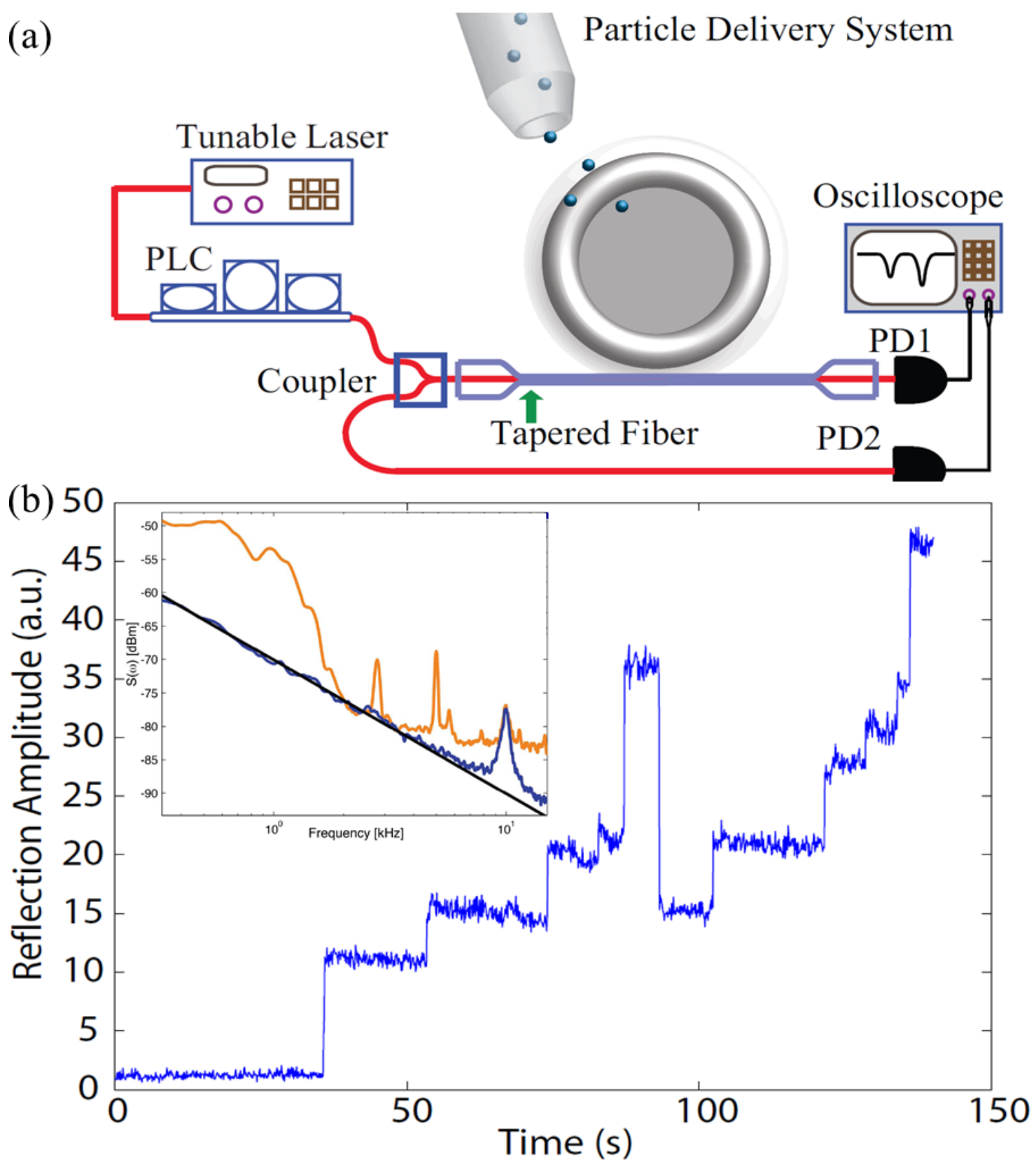

Fig. 7. (a) Experiment setup for back-reflection based nanoparticle detection. (b) The change of reflection signal for the detection of NaCl particles with radii of 20 nm. Inset: Frequency noise comparing for back-scattering spectrum (blue/dark trace) and PDH spectrum (orange/light). Black line: $1/f^2$ fit to back-scatter data. [90], [91]

## G. Optomechanics

The aforementioned sensing mechanisms and techniques are based on the optical spectrum of the microresonator systems. An alternate sensing method is to directly measure the optomechanical spectrum of a microresonator [92]–[98], which has been achieve in both microsphere and microcapillary resonators. [99]–[101] Specifically, Yu *et al.* have developed a sensing mechanism utilizing a cavity optomechanical spring effect, in which single particles/molecules induced optical resonant shift was transferred to the mechanical resonant frequency shift (Fig. 8(a)). [100] Single BSA molecules detection has been achieved with this optomechanical sensing mechanism with a signal-to-noise ratio of 16.8 (Fig. 8(a)). On the other hand, Han *et al.* proposed a microfluidic optomechanical sensor for monitoring flowing particles in a fluid (Fig. 8(b)). [99] Particularly, the sensitivity of this sensing mechanism is highest when the analytes are located at the center of the capillary, which is, far away from the optical mode, ensuring a 100% detection efficiency of the particles flowing in the cavity. Experimentally, single, 3 μm bio-particles was detected in real time. Optomechanical sensing mechanism can realize faster sensing via the ultrafast speed of the real-time electronic spectrum analyzers, which are discussed in Section 5B (time resolution).

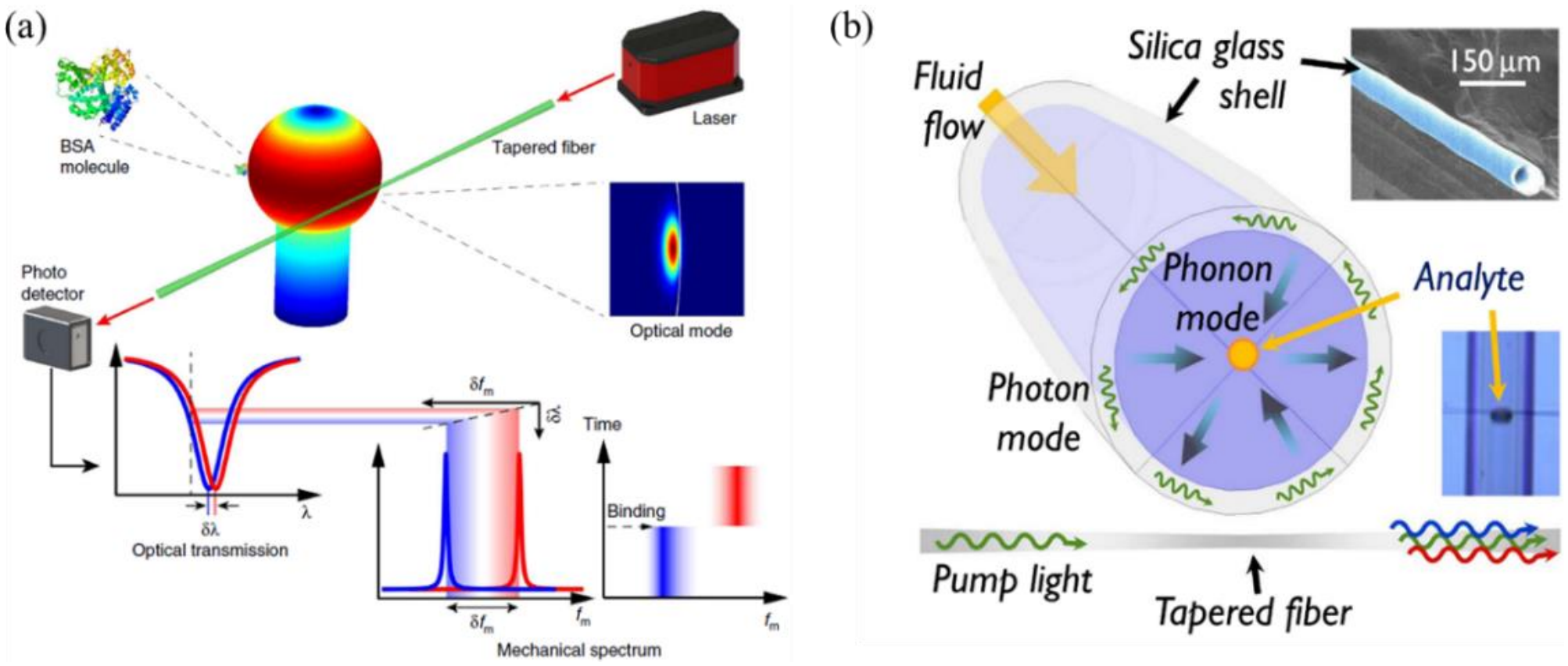

Fig. 8. (a) Diagram of the optomechanical spring sensing mechanism. Both optical and mechanical frequency shifts occur as a result of adsorption of a BSA molecule onto the microsphere surface. [100] (b) Schematic of a cavity optomechanical-fluidic resonator and the principle of optomechanical shift sensing. The phonon mode mediates a long-distance interaction between optical mode and the analyte particles flowing deep witin the fluidic channel. [99]

## IV. Sensing platform

There are many WGM microresonator structures for sensing, including microspheres, microbottles, microbubble/capillaries, microtoroids, microrings, and microdisks, as shown in Fig. 9. In this section, we will be briefly discuss these structure as well as the incorporation of microfluidics with them.

### A. Microsphere

WGM microspheres come in many forms. Early research in WGM resonators primarily studied the emission characteristics of liquid droplets, which are dielectric microspheres. These works will not be covered in this review but an excellent review can be found in Ref. [102] More recent works on sensing using microsphere resonators can be categorized into those that study the emission spectra from dye-doped microsphere, [103] those that directly probe the WGMs of unsupported microspheres made of materials such as barium titanate and PS, [104]–[106] and those that use stem-supported silica microspheres, typically made on the tip of an optical fiber. [19], [107]–[110] Among these, the stem-supported silica microspheres have shown the best performance as sensor, due to the high $Q$ factor that results from the low absorption loss of silica and the low scattering loss from reflow. $Q$ factors as high as $0.9 \times 10^{10}$ has been demonstrated in air, [107] but when used for sensing applications in water, typical $Q$ factor is around $10^5$ to $10^6$. [9], [19], [20], [48]

### B. Microbottle

A bottle microresonator is essentially a cylindrical WGM resonator, in which the cylinder diameter varies along the axial direction, resulting in weak confinement of light in the axial direction. Typically made from standard optical fiber, it consists of a thick portion of silica fiber sandwiched by two thin portions of silica fiber. [111]–[116] The *Q* factor for a bottle microresonator can exceed $10^8$ in air. [111], [112] Bottle resonators have unique caustics in the mode profile, which can conveniently be used for coupling to tapered fiber in an add-drop configuration. [117] Additionally, coupling to WGMs are possible within a large region along the axis dimension of the micro-bottle, making a robust coupling to tapered fiber possible even in an expanding/shrinking matrix such as hydrogel. [118]

### C. Microbubble/Capillary/LCORR

Microbubbles, [119]–[125] and Liquid Core Optical Ring Resonators (LCORRs) [21] represent a class of unique WGM resonators in which the center of the resonator is hollow and can be filled with either liquid or gas. The wall of the capillaries is made thin enough such that WGMs supported in the capillary have significant overlap with the analyte inside the hollow center of the resonator. These resonators have a clear advantage in analyte delivery, especially when used in combination with microfluidic systems. In addition, the flow of gas or liquid inside the capillary does not disturb coupling to the resonator, and makes this type of resonator well-compatible with tapered fiber coupling.

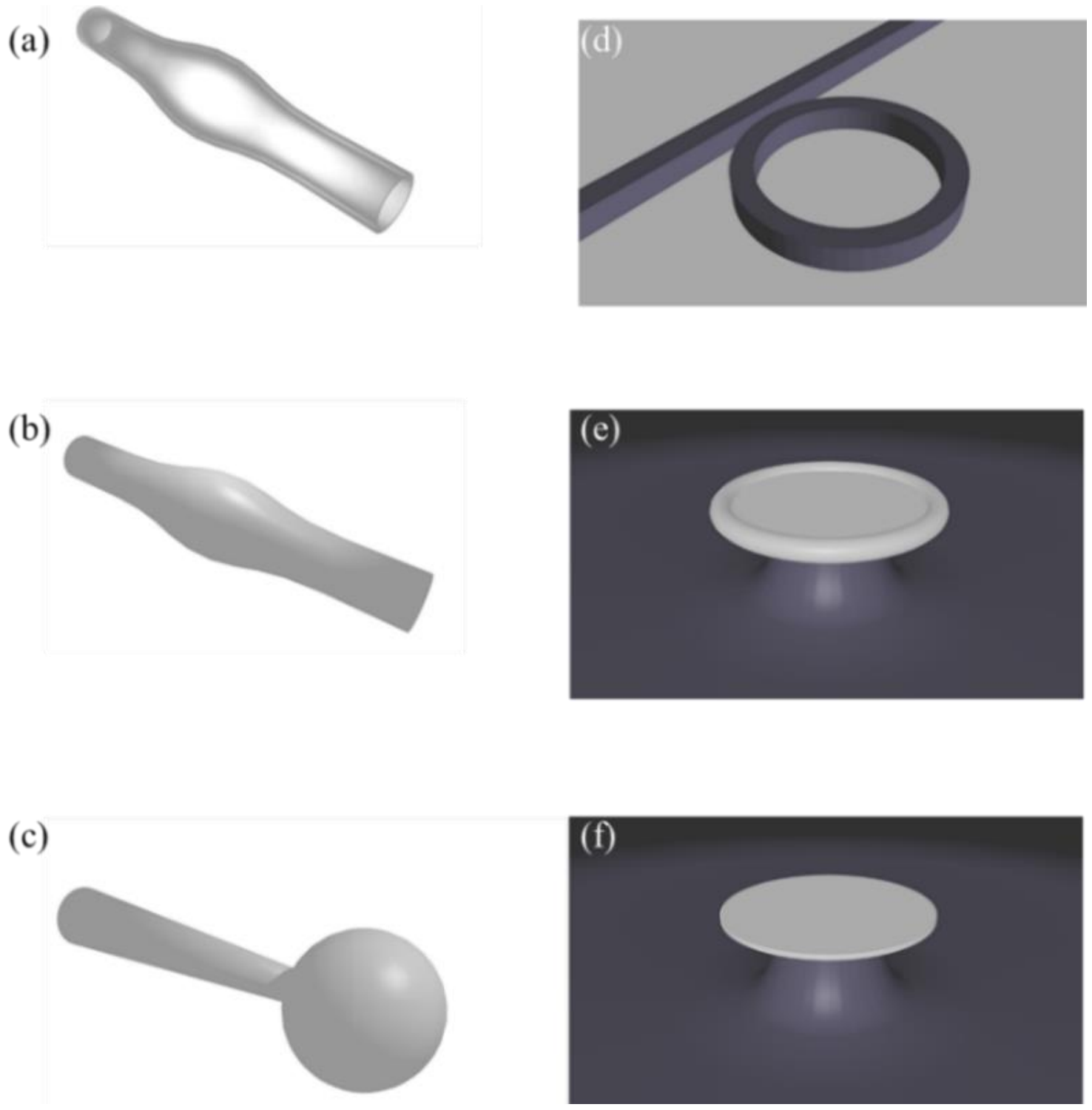

Fig. 9. Illustrations of various resonator geometries: (a) microcapillary, (b) microbottle, (c) microsphere, (d) microring, (e) microtoroid, and (f) microdisk.

## D. Microtoroid

A microtoroid resonator consists of a silica toroid supported by a silicon pillar, fabricated on a silicon wafer. Briefly, the fabrication of a microtoroid involves four steps. [126] First, a photoresist disk is created on thermal oxide grown on top of silicon wafer using photolithography. Second, this disk pattern is transferred to the thermal oxide layer using hydrofluoric acid (HF) etching. Third, xenon difluoride ($XeF_2$) is used for an isotropic, selective etching of silicon. This forms a silica disk on top of a round silicon pillar. Finally, a $CO_2$ laser is used to heat the silica disk, causing the silica at the edge to melt. Due to surface tension, the silica disk shrinks and bulges at the edge, forming a toroidal structure. A microtoroid resonator fabricated this way can achieve $Q$ factor as high as $10^8$. [126] Recently, Knapper *et al.* also developed a method to fabricate all-glass microtoroid that can be reflowed in an oven, which would allow for wafer-scale fabrication of microtoroids. [127]

Microtoroids are fabricated on-chip using lithography, and this allows for an easier integration with other wafer-based technologies and more precise control of device geometry. In particular, in microtoroids, the minor diameter of the torus can be changed independently of the major diameter. This allows for the suppression of many modes that are present in microspheres. [128]–[130] For sensing applications, this is advantageous for observing WGMs that are well separated spectrally. In addition, due to the small minor diameter of the torus optical modes are more tightly confined in a microtoroid compared with a microsphere, and the mode volume of a microtoroid can be smaller than that of a microsphere with equal diameter by about a factor of 5. [131]

## E. Microring

Microrings are made with photolithography and etching, without laser reflow or other high temperature process, making it fully compatible with chip scale integration. Unlike the other types of resonators which are coupled through fiber taper or prism, microrings are almost always coupled through bus waveguide fabricated on the same chip. This allows for a robust coupling that is suitable for integration with microfluidic systems. Different materials have been demonstrated for microring resonator, including silicon-on-insulator, [132], [133] silicon nitride, [134], [135] and polymers. [136]–[138] Ring resonators typically have $Q$ factor on the order of $10^4$ to $10^5$. The relatively low $Q$ factor of microring resonators can be attributed to the surface roughness of the microring due to fabrication and leakage of light into the underlying substrate.

## F. Microdisk

Microdisks are fabricated from lithographically defined disks by wet and dry chemical etching, in a similar manner to microtoroids. The difference between microdisks and microtoroids are in the lack of a reflow process in microdisk resonator. Because of the lack of reflow, the surface roughness on microdisk resonators tends to limit its $Q$ factor, but this also makes microdisks more compatible with on-chip integration. For example, microdisks can be integrated with micromechanical devices to sense force and mechanical motions. [139] In addition, unlike microtoroids, which are almost entirely limited to silica because of the required reflow, microdisks can be made from various materials, making it a popular geometry for the demonstration of new material for WGM resonators. In addition to silica [139], [140] and silicon, [141], [142] microdisk resonators have been demonstrated in silicon nitride, [143] titanium dioxide, [144] silicon carbide, [145] lithium niobite, [146] chalcogenide glass, [147] and polymers. [148]

### G. Microfluidics

The ability to effectively handle fluids with WGM devices pose a significant challenge due to the requirement of a fiber taper or waveguide to couple light into a resonant structure. Initial studies utilizing liquids typically employed microdroplets or small sample chambers. [149] These set-ups, while functional, often suffered from problems due to evaporation or difficulties in incorporating the fiber taper with the fluid cell. Utilizing prism coupling, Vollmer *et al.* has been able to create fluidic cells compatible with ultra-high-*Q* WGM devices, such as microspheres, although these face difficulties in incorporating flow as well as requiring precise coupling to the WGM device itself.

Another solution has been to use on-chip, planar waveguides, and overlying fluidic channels. Bailey and colleagues have utilized Mylar gaskets with Teflon lids to generate fluidic channels over their devices. [14], [150] This technique allows for a rapid means of incorporating fluidic channels onto WGM sensors. Unfortunately, the dimensions of the channels are limited by the ability to process Mylar gaskets. This method also excludes a variety of high-*Q* geometries, such as microtoroids or microspheres, which typically require fiber tapers.

An alternate method of incorporating microfluidics has been to develop WGM devices directly out of components that handle fluid – namely capillaries. One of the first examples of this was the Liquid Core Optical Ring Resonators (LCORR). [21], [151]–[153] This technique was further expanded with microbubble resonators. [119], [120], [154] These devices have been discussed in more detail in Sec. IV C of this review.

Despite a variety of approaches for incorporating microfluidic handling onto WGM devices, there is still a significant need for robust microfluidic handling systems to be integrated with high-*Q* WGM devices. [155] The ability to quickly and reproducibly integrate fluid handling to these highly sensitive devices are critical in their development towards myriad applications.

## V. Sensing Parameters

### A. Sensitivity

The functional sensitivity of the WGM device can be split into two components – the performance of the microresonator itself, and external factors. The performance of a WGM device fundamentally depends on both the construction of the microresonator as well as the sensing modality being employed. For example, in mode shift experiments, a higher *Q* factor allows for a sharper resonance peak, in turn making it easier to detect shifts in the resonance, ultimately improving the sensitivity. In contrast, a high *Q* factor does not contribute to an increased sensitivity markedly in mode broadening based detection. [39] On the other hand, the properties of cavity materials play a key role on some sensors, such as temperature sensing, gas sensing, electric/magnetic field sensing, and pressure/force sensing, which will be discussed in the corresponding section. Each microresonator will have a fundamental limit of detection attributed to its design.

External factors that drastically affect the functional sensitivity of a WGM device include any chemoresponsive elements on the device surface (including capture agents such as antibodies), the medium in which experiments are performed, thermal fluctuations, and other external noise. These factors are largely dependent on the experiment and analyte, making it incredibly difficult to standardize and compare the functional sensitivity of different target analytes.

Regardless of the device engineering and external factors, the functional sensitivity for a microresonator is limited by whichever of the two broad categories provides the poorer sensitivity. A highly sensitive device with a poor capture agent is limited by the capture agent. Similarly, a device with poor inherent sensitivity but with a robust capture agent is limited by the design of the device. Nonetheless, in designing experiments with WGM devices, researchers should be cognizant of both the device design, as well as significant external factors that may ultimately limit.

## B. Time resolution

As mentioned before, the time resolutions of most WGM sensing mechanisms as well as the enhanced techniques are on the order of tens of milliseconds, which are limited by the frequency modulation bandwidth of the laser instead of WGM sensor itself. To improve the time resolution, one can increase the laser sweep speed or utilize techniques that do not require laser scanning. The frequency modulation bandwidth of a tunable laser is typically on the order of kilohertz, corresponding to a time resolution of sub-millisecond. On the other hand, to remove the requirement of frequency sweeping, several techniques have been developed, such as mode locking [85], [86], optomechanics sensing [99], [100], reflection spectrum [90], [91], self-heterodyned microlaser [24], and cavity ring-up spectrum [87]. The time resolution of mode locking sensing technique is about 1 millisecond, which is limited by the low pass filter in the locking system [86]. For the optomechanics mode shift sensing, the time resolution is also around several milliseconds, but is limited by the real-time electrical spectrum analyzer (ESA). The back-scattering reflection spectrum as well as the self-heterodyned microlaser technique could also achieve real-time sensing by locking the frequency of the probe laser to that of the resonant mode. The time resolution is then mainly limited by the data acquisition system and can be on the order of microseconds. The current state-of-the-art time resolution is as short as 16 ns, which is achieve by the cavity ring-up spectroscopy system [87]. Figure 10 briefly summaries both the time resolution and the detection limit of current sensing mechanisms and enhanced techniques.

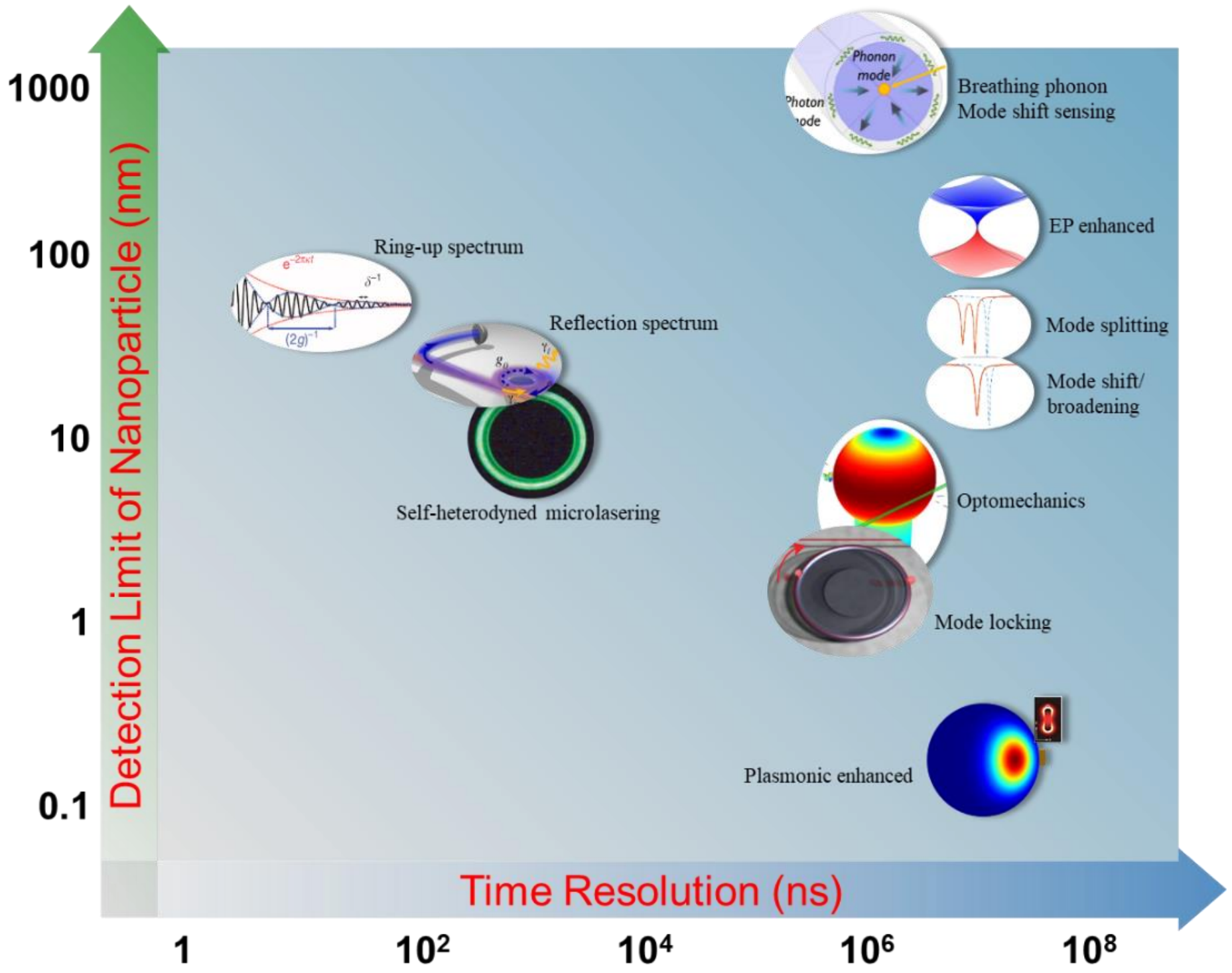

Fig. 10. Time resolution and detection limit of current state-of-the-art sensing mechanisms and enhanced techniques.

## C. Stability

One of the advantages of WGM sensors is their response to any external fields as well as environment changes that influence the refractive index, which could be used for all kinds of sensors. However, on the other hand, this susceptibility to the environment becomes a source of noise in the case a specific sensing signal is desired, which needs to be extracted from these background fluctuations. For example, the detection limits of WGM biosensors based on mode shift are typically limited by all kind of environmental noises, such as the environmental temperature perturbations, mechanical vibrations, and probe laser instability. In contrast, the other two fundamental sensing mechanisms, *i.e.*, mode splitting and broadening, are immune to these environmental background noises due to their intrinsic self-reference property [23], [39]. Mode broadening, for example, is immune to both environmental temperature and mechanical perturbations in that a specially designed PDMS-coated deformed microresonator was used in the experiment. Specifically, the thermo-optic noise induced by both the probe laser and the environmental temperature drift was significantly suppressed due to the reversed thermo-optic coefficients of PDMS and silica. On the other hand, a free-space coupling method was utilized to couple the probe light in and out of the resonator, which removed the extra coupling loss caused by an external near-field coupler. Consequently, the mode linewidth is independent of the coupling positions, as shown in Fig. 11(a). [39]

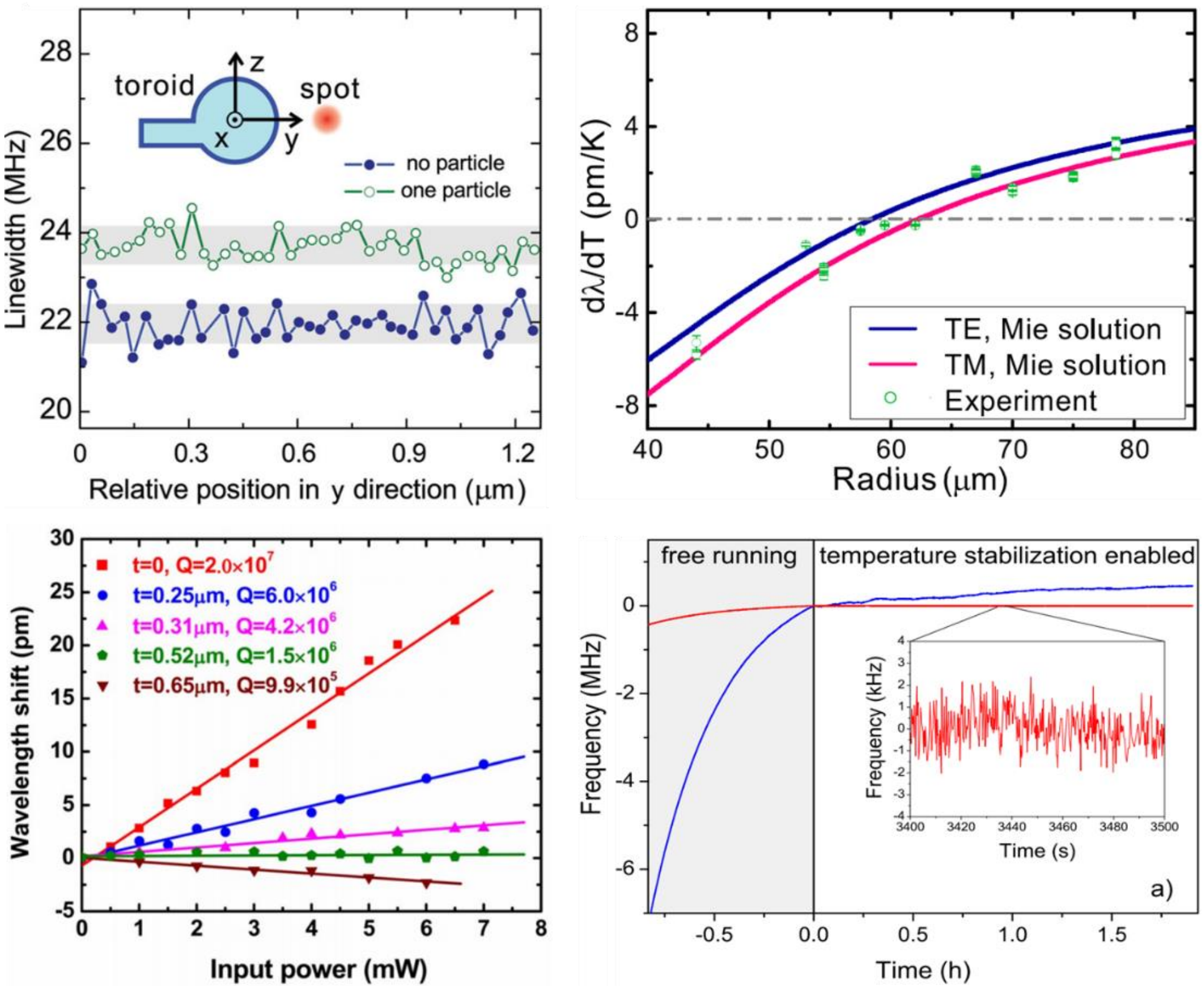

Fig. 11. Techniques for stability in WGM microsensors. (a) Mode linewidth mechanical stability for free-space coupled deformed microresonator. Inset: coordinates definition. [39] (b) Experimental thermal sensitivities (circles) of microsphere resonators of different sizes in the 30% glycerol content in aqueous medium. Solid lines show the theoretical results for both TE (blue) and TM (red) modes. [156] (c) Thermal compensation of a fundamental WGM by coating a particular layer (0.52 μm) of PDMS onto a silica microtoroid. [157] (d) Long-term absolute resonant frequency and dual-mode frequency are shown as the blue and red curves, respectively, in the case of temperature free-running ($t < 0$ h) and stabilized ($t > 0$ h) by the dual-mode technique. Inset: Magnification of the dual-mode frequency fluctuation. [158]

To remove the thermal noise in mode shift, including both laser induced heating [159] and environmental thermal drift, a variety of techniques have been developed. For example, Grudinin *et al.* compensated probe laser scanning induced heating by applying a second stabilization laser sweeping in the opposite direction. [160] He *et al.* demonstrated that a thin layer of PDMS coated onto the surface of the silica microtoroid could eliminate the thermal shift and broadening due to the reversed thermo-optic coefficients induced thermo-optic compensation (Fig. 11(c)). [157] Similar experiments have also been realized in KD-310 glue [161] or quantum dot [162] coated microspheres. In addition, compensation of thermal effect by surrounding medium has also been demonstrated in liquid core microring [151] and microsphere [156] resonators to reduce the WGM's thermal sensitivity (Fig. 11(b)). Furthermore, self-referenced temperature stabilization techniques down to nanokelvin precision have been developed by monitoring the TE and TM modes simultaneously (Fig. 11(d)). [158], [163]–[165]

D. Specificity

The specificity of WGM microresonators largely depends on the chemical composition and functionalization of the device. While the majority of WGM microresonators are composed of silica, the development of structures with myriad of materials has increased the potential chemical functionalizations and capture agents possible. There have been WGM microresonators composed of silicon, PDMS, lithium-niobate, and even silk. The native surface of these devices all influence the amount of non-specific adsorption and fouling that can occur at the surface.

Specificity for WGM microresonators is generally imbued via chemoresponsive elements attached to the surface of the device, as shown in Fig. 12. For chemical species, these include polymer films, brushes, and aptamers. In contrast, for biological molecules, the capture agent of interest is determined by the target antigen. For proteins, antibodies or aptamers are the typical capture agent of interest. For nucleic acids, complementary strands of nucleic acids (cDNAs or LNAs) are typically employed. As mentioned previously, these capture agents not only affect the specificity of measurements but result in inherent limitations to the binding of target molecules at equilibrium.

Another factor that influences the specificity of WGM microresonators are the media in which detection is occurring. For sensing in gases, or neat buffered solutions, the risk of “fouling”, that is non-specific adsorption of molecules to the sensor surface, is minimal. In contrast, measurements in complex media, such a whole blood, run the risk of significant “fouling” and non-specific adsorption.

It should also be noted that for many high sensitivity applications, such as studying the biophysical interactions of these molecules, a specific capture agent is not required.

Another strategy to imbue sensing specificity to WGM microresonators is through the use of combined WGM sensing and Raman spectroscopy. By pumping the Raman scattering resonantly through a WGM, the same molecules that are probed by one of the WGM sensing mechanisms can also be probed by Raman spectroscopy, allowing for WGM resonators to obtain the specific fingerprint from the measured molecules. In addition, resonant pumping provides significantly enhanced Raman signal due to intracavity power buildup. Experimentally, such measurement was done for rhodamine 6G molecules coated on silica microsphere, with a Raman enhancement factor on the order of $10^4$ reported. [166]–[168] However, the use of this technique for biochemical sensing application is yet to be demonstrated.

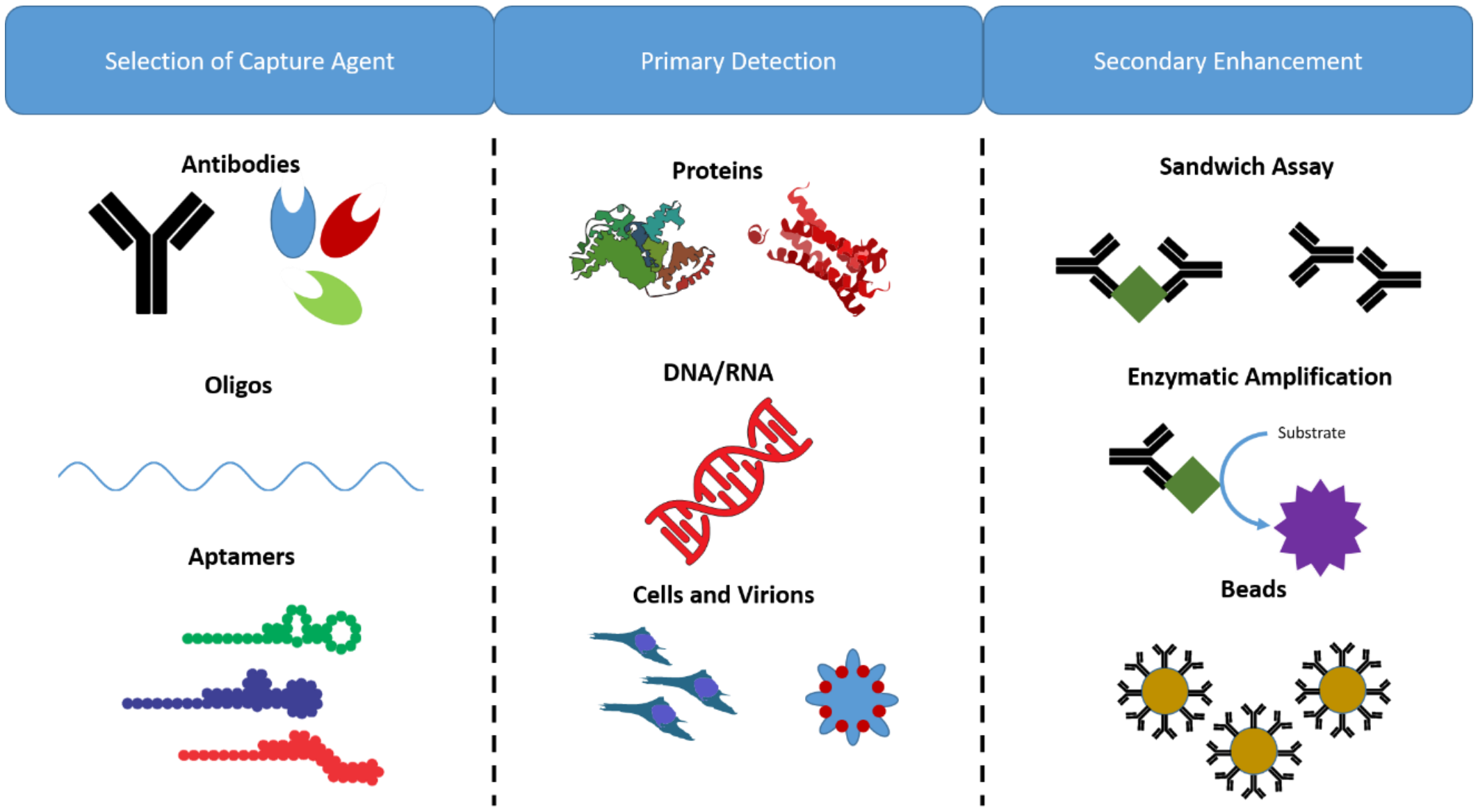

Fig. 12. Schematic detailing the selection of capture agents for biological sensing applications. (Left) Depending on the target molecule of interest, a variety of capture agents can be employed. Following primary detection (middle), additional amplification steps can be pursued to further enhance the sensor response or specificity.

# VI. Sensing Applications

## A. Particle Sensing

The sensing of nanoparticles, typically made of gold, PS and silica in the size range of tens to hundreds of nanometers, represent a convenient model system to study the use of WGM resonators for sensing applications. At the same time, the study of atmospheric aerosol particles and virion particles are important for environmental monitoring and disease control. The detection of single Influenza A virion particles (approximately 50 nm radius) using the mode shift technique was first reported by Vollmer *et al.* in 2008. [9] Since then, virus particles sensing has been demonstrated with mode-splitting, [24] mode-broadening, [39] and in hybrid WGM-plasmonic resonators. [46] The mode-splitting technique is particularly suited for the detection of nanoparticles because the particle-scattering-based mode-splitting signal is unperturbed by environmental perturbations. In addition, real time sizing of the particle size is possible from the measurement of mode-splitting. [23], [38] Decreasing the detection limit of mode-splitting technique is possible by using optical gain through erbium doping [24] or Raman gain. [28], [29] Besides, Zhang et al. recently developed an approach to size single nanoparticle with far-field laser emissions by observing the far-field emission pattern of a pair of coupled deformed resonators. [169], [170] For the mode shift technique, further improvement in signal-to-noise ratio is possible by implementing laser-frequency locking and filtering techniques. [86] These techniques that are developed for nanoparticle sensing can be used for even smaller entities such as single protein and single oligonucleotides, which are covered in the next section. The detection limits of single nanoparticle/molecule for different sensing mechanisms and enhanced techniques are summarized in Fig. 10. Note that sub-nanometer particle (single atomic ion) can be detected using a hybrid gold nanorod-WGM resonator.

## B. Biological Sensing

One area in which WGM's have attracted significant attention is within biological sensing applications. The ability to integrate these devices onto chips, with high sensitivity and low analyte volume requirements, makes them especially appealing for biological applications, in which samples are often limited.

Proteins are the most frequent biomolecule detected with WGM devices, due to their robust clinical roles. Initial studies demonstrated exquisite sensitivity of WGM devices towards proteins, although these were typically performed in neat buffered solutions, and in many cases, without specific capture agents. [104] There have been a number of groups to demonstrate the direct detection of protein in both neat buffered solutions as well as complex media. [150], [171]–[174] In these experiments, a capture agent, typically an antibody or aptamer, is attached to the device surface which provides specificity, as shown in Fig. 13(a). To further enhance sensitivity, as well as specificity in complex media, researchers have also employed a number of secondary enhancement steps including a combination of sandwich assays (Fig. 13(b)), enzymatic reactions, [175] and beads. [174], [176], [177]

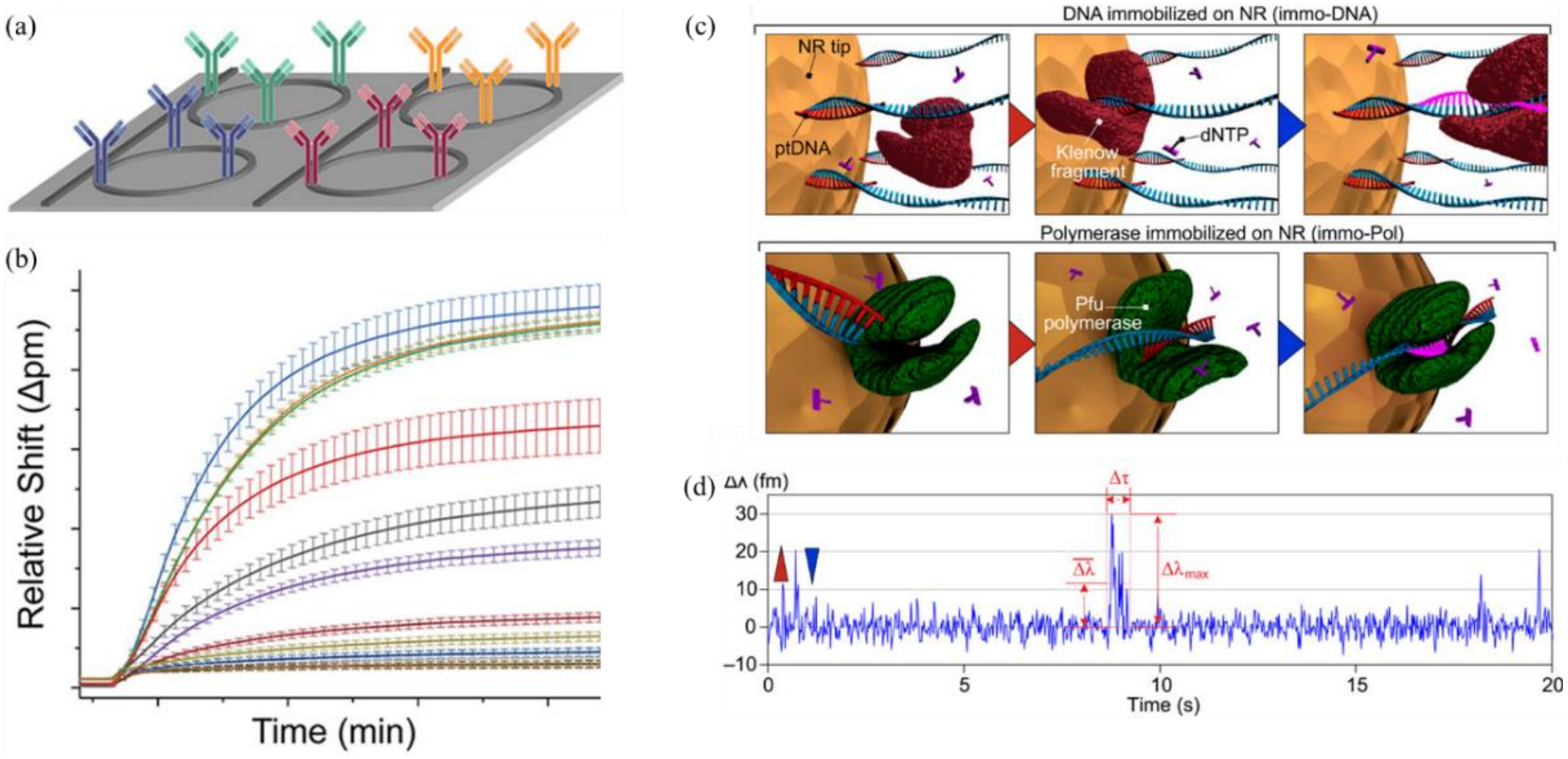

Fig. 13. (a) Schematic diagram of a multiplexed, microring assay for the detection of phosphoproteins. Each individual microring is functionalized with a specific capture protein, specific towards a different target. Spatial separation of the capture probes to different microrings allows for multiplexed measurements. (b) Wavelength shifts corresponding to different concentrations of target protein in a sandwich assay configuration in a multiplexed microring configuration. On the opposite end of the spectrum, (c) diagrams two separate enzymatic activities with DNA and (d) the corresponding resonance wavelength shifts for these single molecule interactions.

Another class of biomolecules that has been extensively studied with WGM devices are nucleic acids, which are conventionally detected via Polymerase-Chain Reaction (PCR) based techniques. Several groups have demonstrated the label-free detection of DNA with WGM devices, including microrings, [178] liquid core optical ring resonators, [179], [180] and microspheres. [181] RNA, which is generally less stable and more difficult to detect with PCR-based techniques, has also been detected using WGM devices. Similar to the detection of proteins, techniques have been employed to further increase the analytical sensitivity for nucleic acids, either through the additional secondary labels and/or enzymatic reactions. [177], [182], [183] Of note, Wu and colleagues developed a novel catalytical network of DNA probes that makes use of sequential probe displacement to amplify both their analytical sensitivity and specificity. [184] Further pushing the limits of sensitivity, Baaske and colleagues demonstrated that with the use of gold nanorods in conjunction with a microsphere resonator, interactions between single strands of nucleic acids could be observed (Figs. 13(c) and 13(d)). [47] Not only did this work demonstrate the exquisite sensitivity of WGM devices towards nucleic acids, but allows for new studies into the fundamental interactions of nucleic acids.

WGM devices have also been used as a method for detecting entire cells and virions. Anderson and colleagues demonstrated the non-specific binding of Helicobacter hepaticus. [185] Ghalia *et al.* used a phage specific protein to capture Staphylcoccus aureus, [186], [187] onto a microdisk. Gohring and Fan were able to detect and subtype human T-cells. Given their relatively small size, traditional optical sensing techniques have difficulties in the detection of viruses. [188] WGM devices have been applied towards the detection of Influenza A, M13, and the Bean pod mottle virus. [9], [24], [38], [189]–[192] While further work is necessary to validate the detection of cells in complex media, these initial studies suggest a potential role for WGM devices as a replacement for traditional biochemical techniques.

## C. Temperature Sensing

While a significant limitation of many WGM devices is thermal drift, many groups have been able to utilize WGM devices as highly sensitive temperature sensors. [114], [165], [193]–[209] Temperature sensing has been demonstrated using silica and silicon based devices. [203], [210] However, the thermo-optic coefficient and thermal expansion coefficient of these materials are positive, making it incredibly difficult to separate the two effects of heating with bare silica/silicon alone. Many WGM devices utilized for temperature sensing are composed of materials that offer a larger negative thermos-optic coefficient, including PDMS, [157], [202], [205], [211], [212] UV-curable adhesives, [112], [203], [213] lithium niobate, [199] and dye-doped photoresists, [214] which give rise to a much higher sensitivity. In addition, materials with large thermal expansion coefficient, such as silk, was also used to fabricate a WGM microresonator thermal sensor. [215]

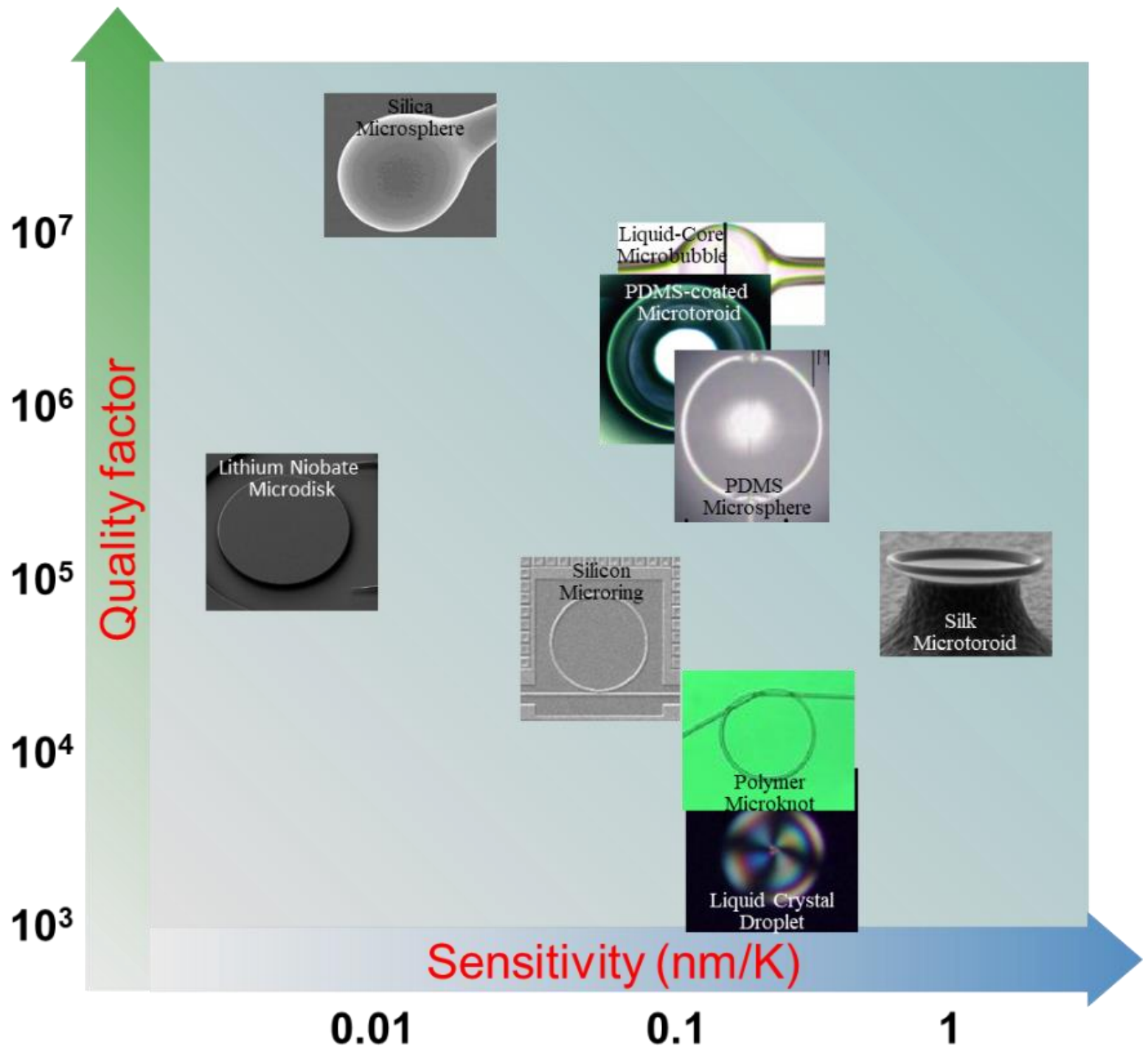

Fig. 14. Quality factor and sensitivity of WGM thermal sensors working on the room temperature.

An alternative experimental design for temperature sensing is the use of microdroplets. The microdroplets themselves can be composed of a variety of materials, including dye-doped cholesteric liquids, nemantic liquid crystals, [197] and oils. [198] The advantage of microdroplet based devices for thermal detection is the ease by which the sensors can be integrated with conventional microfluidics, as well as the ability to utilize materials with relatively high thermal refraction coefficients. Ward and colleagues were also able to utilize thin-shelled, microbubbles filled with air as a modality for temperature sensing. [216] Figure 14 summarizes the sensitivity of some typical WGM thermal sensors working on the room temperature as well as their $Q$ factors. Note that silk microtoroid thermal sensor possesses the highest sensitivity (1.17 nm/K), benefiting from the ultrahigh thermal expansion coefficient of the silk material. [215]

D. Gas Sensing

Another area of sensing in which WGM devices have been applied are for the detection of gases. The most common, and arguably simple, experimental set-up is to coat a WGM device with a chemoresponsive layer specific for the gas of interest. Interactions of the target gas with the polymer layer leads to a change in the refractive index of the layer, which is subsequently detected by the WGM device. The added benefit is that these polymer layers can provide a level of specificity towards target gases. This has been used by myriad groups for the detection of a wide variety of analytes, including ammonia, [217] water, [218] organic compounds, [219] alcohols, [220]–[222] and helium/argon. [223] Gas chromatography has also been coupled with WGM devices, in which a capillary is used as both the medium for the WGM device and separation process. [224] Another approach for the detection of gases involves the use of graphene to generate Brillouin optomechanical modes, which provides an even higher analytical sensitivity. [225]

## E. Magnetic Field Sensing

Ultrahigh-sensitive magnetic field sensors are indispensable components for a wide range of applications, such as geology, archaeology, mineral exploration, medicine, defense, and aerospace. Taking aerospace for example, magnetometers are essential elements for obtain spacecraft attitude coordinates by measuring the geomagnetic field; another example is satellites to study magnetic space explosions. The current state-of-the-art of ultrahigh sensitive magnetometry is achieved by Superconducting Quantum Interference Devices (SQUIDs), which enable detection of single electron spin [226]. However, operation at liquid-He atmosphere temperatures limits the real applications especially for aerospace. Magnetometers capable of room temperature operation offer significant advantages both in terms of operational costs and range of applications. Due to the $1/r^3$ decay of magnetic dipolar fields, sensor size is one of the critical parameters to further improve the sensitivity of sub-femto-tesla magnetometers. Thus, a number of technologies have developed to achieve a higher sensitivity together with a smaller sensor sizes. Among them, the WGM microcavity based optomechanical magnetometer operating in the 100 pT range is a probable candidate to balance the size and the sensitivity. [227]–[233]

Currently, several types of hybrid magnetometers based on WGM have been developed, such as Terfenol-D embed in the pillar of a microtoroid resonator and micro-magnets integrated in soft polymer material surrounding a microtoroid, as shown in Figs. 15(a) and 15(b), respectively. For the Terfenol-D-microtoroid hybrid magnetometer, a 100 $nT/Hz^{1/2}$ range sensitivity with frequencies down to 2 Hz has been achieved. [229], [230] On the other hand, the micro-magnets-polymer-microtoroid hybrid magnetometer possessing a sensitivity of 880 $pT/Hz^{1/2}$ at frequency of 200 Hz has also been demonstrated by Zhu *et al.* [228] Furthermore, Li *et al.* have recently demonstrated a sensitivity improvement technique by probing a Terfenol-D-microtoroid hybrid magnetometer with a phase squeezed light, the noise floor of which is suppressed by about 40%. Accordingly, this technique improves the sensitivity by 50% in the frequency bandwidth from 30 kHz to 45 kHz. [231]

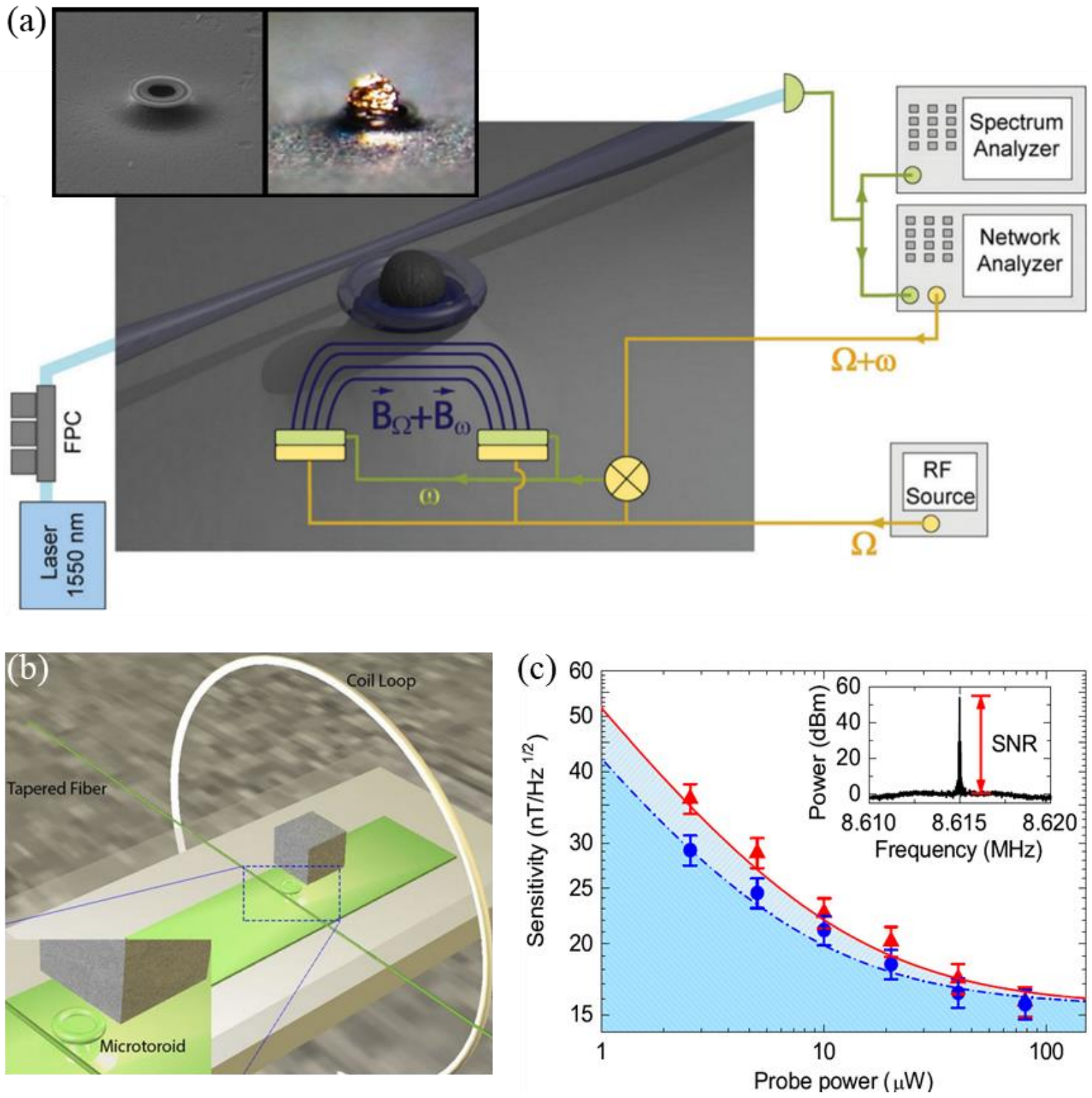

Fig. 15. (a) WGM based magnetometer with Terfenol-D deposition and experimental setup. [229] (b) WGM based magnetometer encapsulated in polymer. The magnet is glued above the resonator chip. [228] (c) Sensitivity of Terfenol-D magnetometer at the frequency of 8.615 MHz, as a function of the probe power. The red triangles and blue circles represent the measured results for coherent and squeezed probes, respectively. The inset shows the power spectrum when the magnetometer is driven at this frequency. [231]

## F. Pressure/Force Sensing

For the detection of pressure and force, there are several configurations utilized by researchers. One configuration involves the use of hollow WGM structures to serve as transducers. Force on the WGM structure itself either leads to a change in the device's shape, or mechanical stresses that are realized as changes in the refractive index of the device, as shown in Fig. 16(a). This technique has been demonstrated with both solid [234] and hollow resonators [235], [236] – the advantage of the latter being that the hollow structure assists in transducing the pressure for measurement (Fig. 16(b)). [237]–[239] Another technique is to immerse the whispering gallery mode device into a transduction medium (such as an elastic polymer) or attached the device directly to a polymer transducer. [239] This technique has been expanded into optomechanical coupling of the WGM devices, in which the resonators are incorporated directly onto mechanical devices.

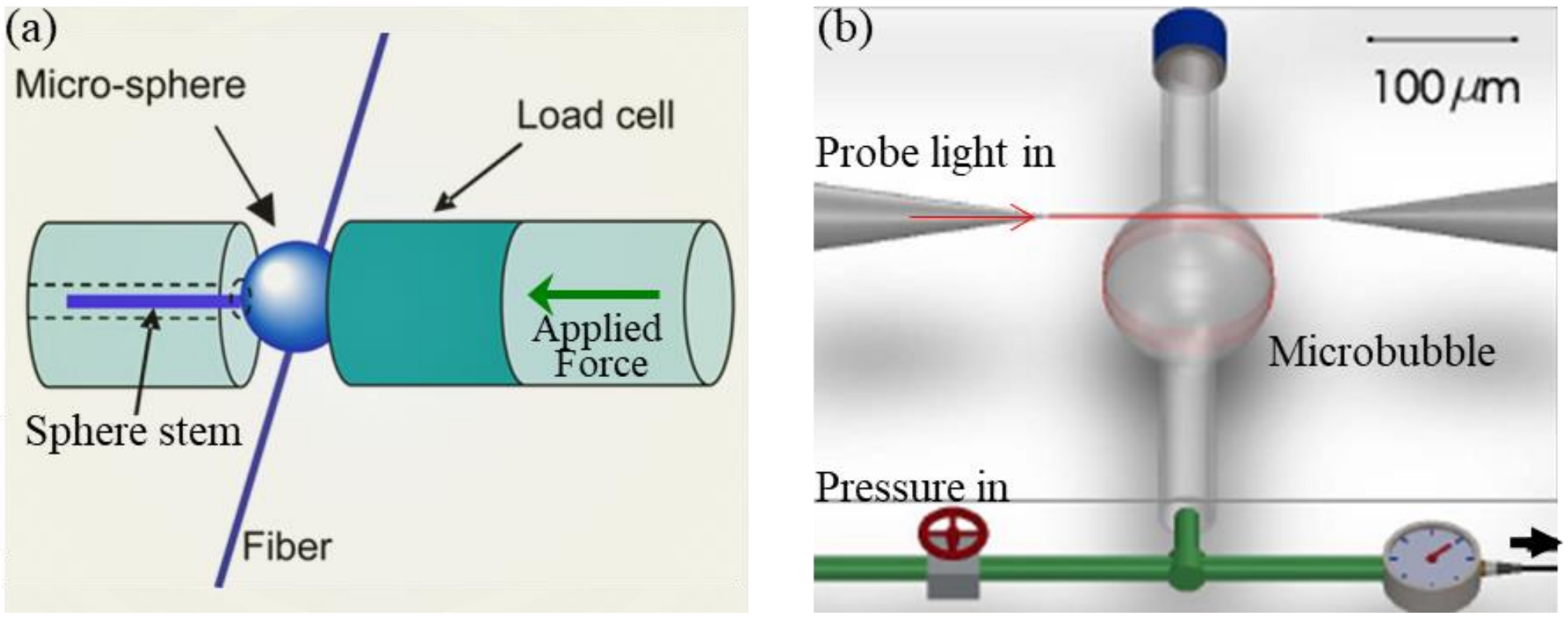

Fig. 16. Schematic of the WGM force/pressure sensors. (a) Microsphere force sensor. [238] (b) Microbubble pressure sensor. [235]

### G. Other Sensing Areas

There is an ever-expanding role of WGM microresonators as sensors. While our review has covered some of the more prevalent areas of sensor development, WGM devices have also been applied as sensors for refractive index [240]–[251], electric fields [252]–[256], gyroscope [234], [257]–[268], humidity [269]–[274], displacement [275]–[282] sensing. We anticipate that the exquisite sensitivity of WGM microresonators will catalyze significant development in these areas moving forward.

## VII. Conclusions and Perspectives

Herein, we have briefly reviewed the mechanisms, methods, structures, technique, parameters, and applications of sensors based on WGM microresonators. Rising from the ultrahigh energy build-up factor of WGMs, all kinds of ultrasensitive sensing experiments have been achieved in the lab, including not only traditional matter sensing, such as particle, gas and bio/chemical sensing, but also field sensing, for example, temperature, electric/magnetic field, and pressure/force sensing, as shown in Fig. 17. Three fundamental sensing mechanisms and several enhanced sensing techniques/mechanisms have been developed in the last two decades, which has been well discussed and summarized. In addition, different kinds of WGM structures as platforms for sensing as well as microfluidics technique are presented. Some sensing parameters, such as sensitivity, time resolution, stability and specificity, have been discussed for some sensing techniques/mechanisms.

Looking ahead, there are still all kinds of challenges and potential directions for WGM microresonator sensing. First is the development of new hybrid materials and structures for the targeted detection of analytes or applications. For example, new material composites can be developed for specific analytes or applications; researchers could select the material their device is composed of based on the target gas to be detected.

Second, there is still significant progress to be made in the development of WGM microresonator structures, such as deformed microresonator [283]–[299], endoscopic sensing probes [300], [301], and WGM sensors in chip-based microfluidics channels. Not only will these lead to further improvement in device sensitivity, but also allow for the detection of analytes that current techniques are currently unable to.

Third, there is an increased need for the integration and miniaturization of the entire WGM sensing systems into single devices. [302] The ability to integrate the laser source, sensor, detectors, and data processing hardware into a single-device will be integral in the adaption of these devices for myriad applications, including those in the field and within the medical realm.

Fourth, we envision the development of techniques that allow for the undirected and *de novo* identification and detection of analytes. This would extend the application of WGM microresonators as a sensor into a new, highly sensitive discovery tool.

Fifth, the development of even more enhancement techniques for sensing purposes. As our understanding of the fundamental physics and phenomena of these devices improves, new and innovative methods to improve their optical performance will follow.

Finally, the scalability of these devices, coupled with the ease by which they can be integrated with other electronics, makes them amenable for the development of network-based sensing. Entire networks of WGM microresonators can be constructed to provide highly multiplexed and sensitivity ensemble measurements.

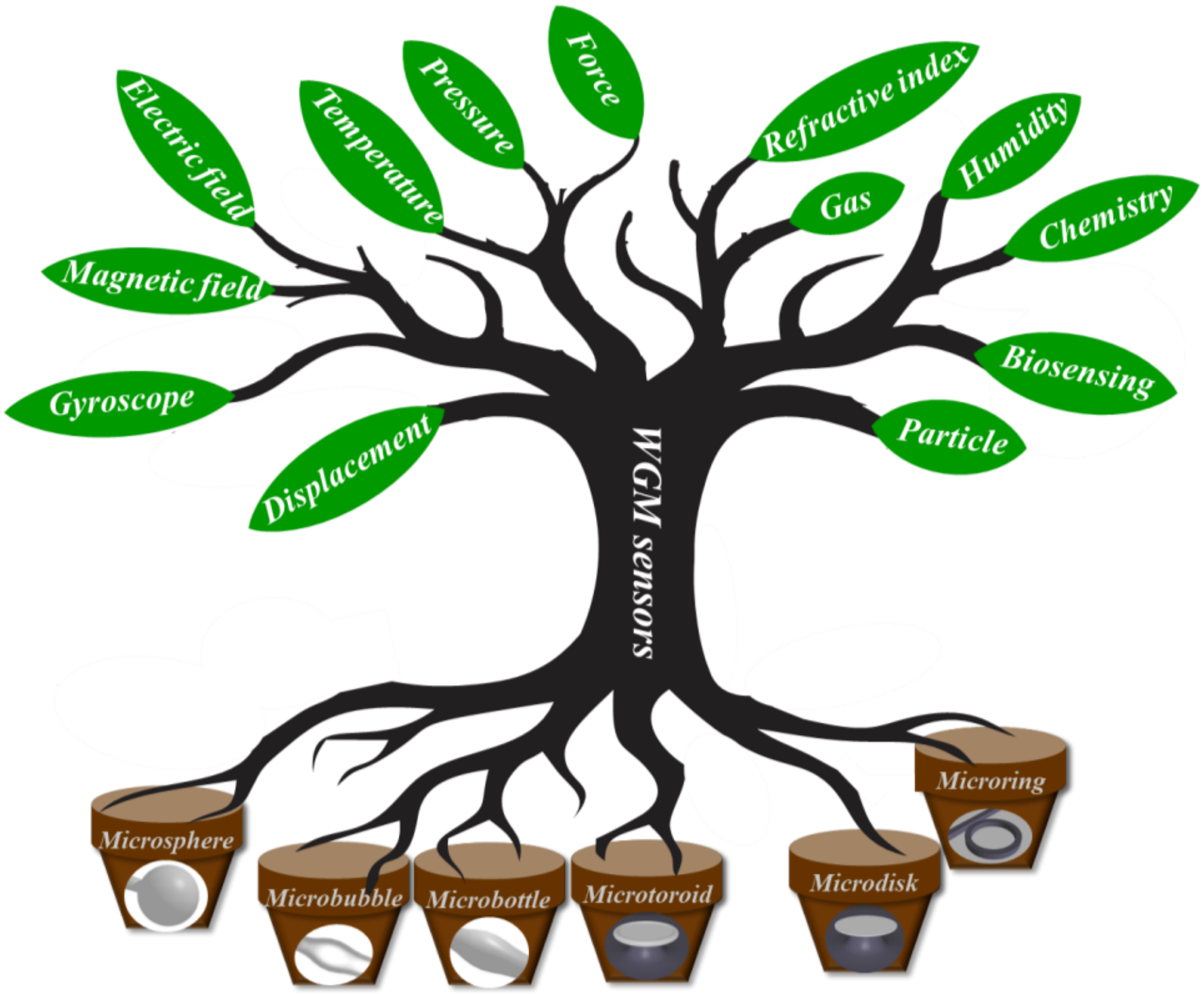

Fig. 17. Representation of the WGM resonator structures and fields of study.